# Explicit Bounds for Entropy Concentration under Linear Constraints

KOSTAS N. OIKONOMOU

AT&T Labs Research
Middletown, NJ 07748 U.S.A.

*Email:* ko@research.att.com

PETER D. GRÜNWALD

CWI
P.O. Box 94079, NL-1090 GB Amsterdam, Netherlands

*Email:* Peter.Grunwald@cwi.nl

## Abstract

Consider the set of all sequences of $n$ outcomes, each taking one of $m$ values, whose frequency vectors satisfy a set of linear constraints. If $m$ is fixed while $n$ increases, most sequences that satisfy the constraints result in frequency vectors whose entropy approaches that of the maximum entropy vector satisfying the constraints. This well-known "entropy concentration" phenomenon underlies the maximum entropy method.

Existing proofs of the concentration phenomenon are based on limits or asymptotics and unrealistically assume that constraints hold precisely, supporting maximum entropy inference more in principle than in practice. We present, for the first time, non-asymptotic, explicit lower bounds on $n$ for a number of variants of the concentration result to hold to any prescribed accuracies, with the constraints holding up to any specified tolerance, taking into account the fact that allocations of discrete units can satisfy constraints only approximately. Again unlike earlier results, we measure concentration not by deviation from the maximum entropy value, but by the $\ell_1$ and $\ell_2$ distances from the maximum entropy-achieving frequency vector. One of our results holds independently of the alphabet size $m$ and is based on a novel proof technique using the multi-dimensional Berry-Esseen theorem. We illustrate and compare our results using various detailed examples.

# Table of contents

# 1   Introduction

The phenomenon of entropy concentration, or "concentration of distributions at entropy maxima" as it was originally called by E.T. Jaynes [Jay83], is significant for a number of reasons. From a mathematical standpoint, it establishes a tight connection between counting elements of constrained sets and maximizing the Shannon entropy. From the viewpoint of probability theory, it underlies the conditional limit theorem for uniform distributions. With respect to inference in probability theory, it may be viewed as powerful support for the principle of maximum entropy (MaxEnt). From the viewpoint of coding theory, entropy concentration directly implies bounds on the minimax codelength in a certain data compression task [Gr1], [Gr8]. Finally, recent work is establishing connections between maximum entropy distributions and estimates for the cardinality of large polyhedral sets ("approximate counting", [BH10], [SV13]).

Despite the fact that the concentration phenomenon has been known for a long time, its status so far has been essentially that of a limit theorem. Our aim here is to bring the concentration phenomenon much closer to the realm of practical application. To do this, we replace all asymptotic considerations with explicit, finite bounds, depending on parameters specifying the desired degree of concentration, the set of admissible solutions besides the one of maximum entropy, and the tolerances to which the constraints have to be satisfied. The main new ingredients are that (a) our results are non-asymptotic, and (b) they allow tolerances on the constraints, a very natural but hitherto unexplored setting. A side-effect of allowing tolerances is that (c) the results are valid *for all n* beyond a certain point, not only for special values of $n$ (see §2.2), also an issue which has not been previously addressed. Finally, (d), we measure concentration not by deviation from the maximum entropy value, as is traditionally done, but by a more intuitive metric, the *norm* of the difference from the maximum entropy solution.

We note that simple non-asymptotic 'reference' results, which we present as Theorems 3.4 and 3.5 in §3.2, can be easily derived using the method of types [CT06], and may therefore not be surprising to information theorists. Here we present two types of new results. Our results of the first type, Theorems 3.14 and 3.15, and the related Theorem 3.17 and Corollary 3.18, give much better bounds than the reference results. They are still similar to the reference results in that, for reasonable values of the tolerances, the results become nontrivial only if the alphabet size $m$ is much smaller than the sample size $n$. However, our second type of result, Theorems 4.1 and 4.4, while also non-asymptotic and including tolerances, can, if the constraints are only equalities, deal with *arbitrarily large m*. Thus, apart from (a)-(d) above, another main ingredient of our paper is, (e), to provide, for the first time, a non-asymptotic result that deals well with large $m$. A first large-$m$ result was established by [Gr1], [Gr8], but using a technique based on local central limit theorems, unsuitable for non-asymptotic, finite-sample bounds. Here we do manage to get such bounds using the multidimensional Berry-Esseen theorem [Ben03], [CF11]. These bounds depend on the *number of (equality) constraints*, and so, in some sense, are dual to the bounds that depend on $m$, the number of 'variables'. When the number of variables is small the first type of result, Theorem 3.14 etc., generally leads to much better bounds.

## The setting

Our prototype is a process which is repeated $n$ times and each repetition has $m$ possible outcomes. For concreteness we will think of assigning $n$ balls to $m$ labelled bins, where each bin can hold any number of balls. The first ball can go into any bin, the second ball can go into any bin, ..., and the $n$th ball can go into any bin. Each assignment or allocation is thus a *sequence of n bin labels* and results in some number $\nu_1$ of balls in bin 1, $\nu_2$ in bin 2, etc., where the $\nu_i$ are $\geqslant 0$ and sum to $n$. There are $m^n$ possible assignments in all, and many of them can lead to the same *count* vector $\nu = (\nu_1, ..., \nu_m)$. We refer to these assignments as the *realizations* of the count vector and denote the number of realizations of $\nu$ by $\#\nu$.

The arrangement of $n$ balls into $m$ bins can represent the construction of any discrete object consisting of $m$ distinguishable parts out of $n$ identical units. So if the balls represent pixels of an image, the attributes of color and (suitably discretized) intensity are ascribed to the bins to which the pixels are assigned. Then the count vector is thought of as a 2-dimensional matrix with rows labelled by intensity and columns by color. In another situation the balls could represent packets in a communications network with attributes of origin, destination, size, and timestamp, and so on. Now consider imposing constraints $\mathcal{C}$ on the allowable assignments to the bins, expressed as a set of *linear* relations on the elements of the *frequency* vector $f = (\nu_1/n, ..., \nu_m/n)$ corresponding to the counts $\nu$. E.g. $5\,f_1 - 17.4\,f_2 \geqslant 0.131$, $3\,f_5 - 4\,f_7 + f_{11} = 0.123$, $f_{12} \leqslant f_{15}$, etc. With $m$ given and fixed, as $n$ grows, the frequency vectors of more and more of the assignments that satisfy the constraints will have entropy closer and closer to that of a particular $m$-vector $\varphi^*$, the vector of *maximum entropy* $H^*$ subject to the constraints $\mathcal{C}$. (We denote this vector by $\varphi^*$, as opposed to $f^*$, to emphasize that its entries are, in general, not rational.) Thus for large $n$, the vast majority of all possible constructions that accord with the constraints will finally yield something close to, or concentrated around, the maximum entropy vector.

Several variations of this result are known: the original is E.T. Jaynes's "entropy concentration theorem" [Jay82], [Jay83]; a related result in the information theory literature is the "conditional limit theorem" [CT06], originally proved in [vCC81], and extended in [Csi84]; more recently we have the "strong entropy concentration" results of Grünwald [Gr1], [Gr8][1.1]. All of these results involve limits or asymptotics in one way or another, i.e. in the statement "given an $\varepsilon > 0$ and an $\eta > 0$, there is a $N(\varepsilon, \eta)$ such that for all $n \geqslant N(\varepsilon, \eta)$ the fraction of assignments that satisfy $\mathcal{C}$ and have a frequency vector with entropy within $\eta$ of $H^*$ is at least $1 - \varepsilon$", one or more of the quantities $\varepsilon$, $\eta$, or $N$ is not given explicitly. (To prevent a possible point of confusion, here we are talking about the *number of ways* of constructing a particular object satisfying the constraints, not about the *number of objects* that satisfy them; the latter number is quite a different matter.)

In our balls-and-bins setting there is *no randomness*, and there are *no probabilities* anywhere: we are simply counting all the possibilities. This fully discrete, constructive, counting setting is unlike the setting in which the maximum entropy (MaxEnt) principle is most often applied, the derivation of real-valued probability distributions from probabilistic information. Nevertheless, by a suitable interpretation of the balls, the object we end up constructing in our setting *can be* a (discrete) probability distribution with rational entries; this is illustrated in §3.7.

To re-emphasize the point that we are simply enumerating all the possibilities, Fig. 1.1 depicts the $3^5$ possible assignments of just 5 balls to 3 bins, and the beginnings of concentration around the maximum entropy value and vector, even in this very small, simple case.

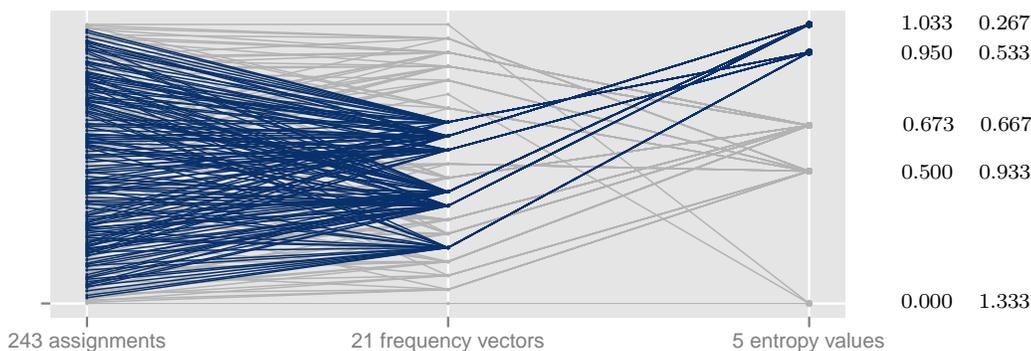

**Figure 1.1.** The 6 frequency vectors with entropies closest to $H^* = 1.099$, and smallest deviations from $\varphi^* = (1/3, 1/3, 1/3)$ in $\ell_1$ norm, have more than half of all the realizations. The correlation between the entropy (first) and norm (second) values is due to the uniformity of $\varphi^*$.

---

[1.1]. All these results make similar statements about similar things, but it is not our purpose here to enter into a detailed comparison.

Our purpose is to construct an $m$-part discrete object of maximum entropy (equivalently, with the largest number of realizations) satisfying the constraints, and to compare it with other 'similar' or 'nearby' objects that also satisfy the constraints. In the existing literature it is customary to compare objects, represented by their frequency vectors, by comparing the entropies of the vectors. We suggest in §3.2 that a measure more suitable for this purpose is the *norm* of the difference of the frequency vectors; both the entropy and norm metrics are shown in Fig. 1.1. We will use a tolerance $\vartheta$ to measure the allowable deviation in norm from the MaxEnt vector $\varphi^*$. Another issue is that some constraints, e.g. $f_1 + f_2 = 1/137$, are not satisfiable by rational numbers with denominator $n$ unless this $n$ is a multiple of the r.h.s., here 137. This would make it impossible to state that a concentration result holds for *all* $n$ beyond a certain point. To make such statements possible we introduce another tolerance, $\delta$, on the satisfaction of the constraints. This tolerance has other interpretations as well: (a) it can reflect uncertainty in the values of the constraints; (b) it allows us to formalize that, due to natural variation, constraints derived from past observations will only hold approximately, up to a certain $\delta > 0$, on future observations about which one wants to make inferences (see Remark 4.3 in §4).

Table 1.1 lists the tolerances involved in our statement *EC* of the entropy concentration phenomenon below.

| | |
|---|---|
| $\delta$: | relative tolerance in satisfying the constraints |
| $\varepsilon$: | concentration tolerance, on ratios of numbers of realizations |
| $\vartheta$: | absolute tolerance in deviation from the MaxEnt vector $\varphi^*$ |

**Table 1.1.** Tolerances for the entropy concentration results.

We have already commented on $\delta$. The tolerance $\varepsilon$ restricts *how often* we can deviate from the most likely object, and the tolerance $\vartheta$ specifies *how much* we can deviate. It turns out that $\delta$ and $\vartheta$ cannot be chosen independently of each other: as we discuss in §2.3, §3.2, and also in §4.1, $\delta$ *cannot be too large* relative to $\vartheta$ or concentration may no longer obtain.

We present three pairs of theorems. For given positive tolerances $\delta$, $\varepsilon$, and $\vartheta$, and number of units or "sample size" $n$, the first theorem in each pair (Theorem 3.4, 3.14, and 4.1) provides an explicit bound on the fraction of assignments that satisfy the constraints $\mathcal{C}$ to accuracy $\delta$ and have a frequency vector no farther than $\vartheta$ from $\varphi^*$ in norm. The second theorem in each pair, Theorem 3.5, 3.15, and 4.4, establishes a statement of the form:

*EC:*       given positive tolerances $\delta, \varepsilon$, and $\vartheta$, as described in Table 1.1, provide an $N(\delta, \varepsilon, \vartheta)$ such that for all $n \geqslant N$, the fraction of assignments that satisfy the constraints $\mathcal{C}$ to accuracy $\delta$ and have a frequency vector no farther than $\vartheta$ from $\varphi^*$ in norm, is at least $1 - \varepsilon$.

The emphasis of these latter results is on minimizing $N$ rather than on obtaining tidy closed forms for it. Therefore the results lack analytical elegance, and read more like algorithms for computing $N$. As already indicated in the first paragraph, the first pair of theorems are a simple reference result; the second pair concerns a highly optimized result in which the number of bins (or alphabet size) $m$ occurs, and the third pair concerns results in which it does not.

## Relevance and related work

The mathematical contribution of our work is that we tighten the connection between entropy maximization and counting elements of constrained sets of $n$ outcomes, each chosen from a set of $m$ elements, giving a connection for arbitrary $n$, and even for arbitrarily large $m > n$. This can be directly reinterpreted as a *precise* version of the conditional limit theorem relative to the uniform distribution (for all details about this interpretation we refer to [CT06], [vCC81], [Csi84] and [Gr8]), and also as providing bounds on the number of bits needed to code data satisfying some constraints with a worst-case optimal code (see 'the empirical coding game' described by [Gr8] for details). If we regard our assignments as resulting from a uniform probability distribution over the $m$ bins, the concentration around the maximum entropy p.d. can be viewed as a statement in the area known as "concentration of measure" (e.g. [DP09], [BLM13], [RS14]). Whereas the usual concentration of measure occurs around the mean or median of a quantity, in our case the measure, be it probability or number of realizations, concentrates around the *maximum entropy* value or probability distribution.

Beyond mathematics, we note that the concentration phenomenon provides support for the principle of maximum entropy, MaxEnt, in inductive inference. Two aspects of the application of this principle are still the subject of debate: the interpretation of the constraints, and its compatibility with Bayesian conditioning or updating. We do not propose to enter this debate here, and of the extensive literature mention only [Gr0] for an overview of 'safe', 'risky' and 'silly' applications of MaxEnt, [Uff96] and [Wil11] which cover a large set of common references, and [Cat12]. We also point out that in our fully discrete setting, devoid even of probabilities, the constraints have a particularly clear interpretation. In the final analysis, in some cases the application of MaxEnt can certainly be justified, and in those cases the concentration phenomenon plays a large role in the justification. However, previous statements of entropy concentration were incomplete as they were always asymptotic (with no clue as to the size of hidden constants), they assumed constraints to hold precisely without any tolerances, which in practice is often unrealistic, and were valid only for those special values of $n$ for which the constraints were satisfied with exactness. We improve these earlier statements by giving non-asymptotic bounds, and allowing tolerances in the constraints, which removes any restrictions to 'special' $n$.

Concerning other justifications of maximum entropy inference not involving probabilities, *axiomatic* justifications independent of probabilities, but under equality constraints only, were given by Csiszár in [Csi91] and [Csi96], and also by Skilling in [Ski89].

We also mention that there is a very large variety of applications of MaxEnt. See the overview [KK92], as well as [Oik12], which has a concentration flavor. Both foundational issues and applications are an active research area, as evidenced by the conference proceedings [ME98] and [MEnt].

One very recent application whose relation to our work deserves some elucidation is *approximate counting* [BH10], [SV13]. In particular, [SV13] also deals with counting elements of discrete sets and distributions that maximize entropies up to some tolerance, and may thus seem closely related–but the actual problems adressed are quite different. The authors consider the problem of efficiently counting the elements of a subset $\mathcal{M} \subset \{0,1\}^m$ when one has access to an oracle that can calculate maximum entropy distributions up to a given tolerance over this subset. To be more precise, let $\theta = (\theta_1, ..., \theta_m)$ be an element of the convex hull of $\mathcal{M}$. Let $C_\theta$ denote the set of distributions on $\mathcal{M}$ satisfying the constraint $\Pr(X_1 = \theta_1), ..., \Pr(X_m = \theta_m)$, where $X_j$ denotes the $j$th bit in the realized outcome of $\mathcal{M}$. Then for any $\theta$ satisfying some mild regularity conditions, when given the vector $\theta$ and a tolerance similar to our $\vartheta$, the oracle outputs a distribution in $C_\theta$ that has entropy within $g(\theta)$ of the maximum entropy distribution in $C_\theta$, where $g$ is some function such that $\lim_{x \downarrow 0} g(x) = 0$. Broadly speaking, [SV13] show that given such an oracle, one can efficiently compute the number of elements of $\mathcal{M}$ up to a factor $1 + \gamma$ in time polynomial in $m$ and $1/\gamma$. They also provide a converse to this statement, showing that given an oracle that can approximately count, one can efficiently calculate maximum entropy distributions for any given constraint vector $\theta$ up to some tolerance. These two results taken together show that, with constraints given in terms of marginals, computationally *efficient* approximate counting is equivalent to computationally *efficient* approximate maximum entropy determination. This is quite different from our results, which have nothing to say about computational efficiency, but instead have an extra variable $n$, the number of outcomes, not present in [SV13], and give precise bounds on the *fraction of sequences* of length $n$ that satisfy given constraints up to a certain tolerance and have frequency vectors within a certain tolerance of the maximum entropy *vector*.

## Summary

In §2 we describe some details about the constraints and the tolerances on them: why tolerances on equality constraints are necessary, how we handle elements of the solution that are forced to 0 by the constraints, assumptions about linear independence, and the effect of introducing tolerances on the constraints on the p.d. that maximizes the entropy, as well as on the resulting value of the maximum entropy. §3 begins with the two benchmark results Theorem 3.4 and 3.5 which serve as a reference for the more sophisticated results that follow. Next we present our first three main results, significant improvements over the 'reference' results at the expense of additional complexity. The first two results concern the concentration on a *set* of vectors around the MaxEnt $\varphi^*$. The third concerns concentration at *the single* discrete object (rational vector) $f^*$ derived from the continuous $\varphi^*$. This has implications on the arising of the uniform probability distribution as the "most preferred" distribution in the absence of any information. All of these results are illustrated by a number of examples. In §4 we present our second pair of main results on concentration. They apply when the constraints consist only of equalities[1,2], but can yield significantly smaller $N$ than the general results of §3 in large-$m$ problems. The $N$ of §4 does not depend on $m$, but on the number of the constraints, so $m$ can even be taken as infinite. These are also combinatorial results, but they are established indirectly, via probabilistic methods, unlike the results of §3 which rest only on combinatorial arguments. We again give a number of examples for illustration.

All our results are interpretable as assignments of possibly indistinguishable balls to distinguishable bins, and therefore show how the discrete object/count vector $\nu^*$ with the desired properties can be actually constructed.

All of the proofs are collected in Appendices.

## 2 Constraints and tolerances

We discuss certain high-level assumptions that are convenient, such as that the constraints of the problem are linearly-independent and that we eliminate all zero elements in the solution from consideration, and some that are necessary, namely that the constraints are subject to tolerances, otherwise it is impossible for the concentration phenomenon to hold for all $n$ beyond a certain point.

### 2.1 Structure of constraints and zeros in the solution

Let $A^E$ be a real $\ell_E \times m$ matrix, $A^I$ a real $\ell_I \times m$ matrix, and $b^E, b^I$ $\ell_E$- and $\ell_I$-column vectors. We consider the problem of maximizing the entropy of a real $m$-vector $x$ subject to linear equality and inequality constraints:

$$\max_{x \in \mathbb{R}^m} H(x) = -\sum_i x_i \ln x_i \qquad \text{subject to} \quad A^E x = b^E, \ A^I x \preccurlyeq b^I, \ \textstyle\sum_i x_i = 1, \ x \succcurlyeq 0, \qquad (2.1)$$

---

1.2. A certain type of inequality is also permissible.

where $\preccurlyeq$ denotes component-wise $\leqslant$. The matrices $A^E, A^I$ represent the *structure* of the constraints, and the vectors $b^E, b^I$ the *data*, or values of the constraints. We will assume that the set of constraints in (2.1) is satisfiable. Then (2.1) is a feasible strictly concave maximization problem, and has a unique optimal solution $\varphi^*$ (see e.g. [BV04], [HUL96]).

Some elements in the solution of (2.1) may turn out to be 0. Some of these may be explicitly set to 0 by the constraints, e.g. by a constraint such as $x_{15} = 0$, and some may be implicitly forced to 0 by the constraints, e.g. given $x_1 + 2\,x_2 = 0.6$, $x_2 = 0.3$, it must be that $x_1 = 0$.

We impose the following requirements on the constraints:

1. We assume that once the entropy maximization problem is solved, all explicit and implicit zeros in the solution are *removed from consideration*, both from the MaxEnt vector itself, and from the matrices and vectors specifying the constraints[2.1]. So in the sequel we will assume that all elements of $\varphi^*$ are strictly positive; $m$ will be the number of these elements, as well as the dimension of $x$ when we refer to the constraints (2.1). The removal of the 0s and adjustment of $m$ is not strictly necessary for the development that follows, but it is a great convenience for the derivations in both §3 and §4.

2. As another convenience, in §4 we assume that the constraints are linearly independent, with one exception. The exception is two-sided inequalities in $A^I$, that is constraints where a particular linear combination of the variables is subject to both lower and upper bounds. Two-sided inequalities become important in §4.3.

We will not be concerned here with how to solve the entropy maximization problem (2.1) numerically (for that see [FRT97], [BV04]), but will assume that the solution $\varphi^*$ is available and is as accurate as necessary. When we do calculate $\varphi^*$ numerically, we use the CVXOPT convex optimization package, [ADV].

## 2.2  Tolerances on constraints

When dealing with a discrete problem, some care is required in connection with equalities in the constraints (2.1), whether they are explicit or implied by inequalities. For example, suppose one constraint is $f_1 + f_2 = 1\,/\,139$. This is not satisfiable by rational numbers with denominator $n$ unless $n$ is a multiple of 139, rendering the statement $EC$ in §1 impossible with $\delta = 0$. If there are many such constraints which must be satisfied exactly, the statement $EC$ with $\delta = 0$ holds under circumstances so special as to render it practically useless for any application. Similar problems with equalities arise because it is convenient to express constraints with coefficients over the real numbers as opposed to over the integers or rationals, and again we want the rational frequency vectors to satisfy these constraints for all $n$ beyond a certain point. Fortunately, when $n$ is large, it is often perfectly acceptable if the equalities are satisfied only to a good approximation, and in fact the same goes for the inequalities. Further, it may be that there is some uncertainty in the precise values $b^E, b^I$ of the constraints. For all of the above reasons we will assume that the constraints need to be satisfied *only approximately*, to within a tolerance $\delta$.

Accordingly we define the set of $m$-vectors $x$ that satisfy the constraints in (2.1), other than non-negativity and normalization, with a relative accuracy or *tolerance* $\delta \geqslant 0$:

$$\mathcal{C}(\delta) \triangleq \{x \in \mathbb{R}^m : b^E - \delta\,\beta^E \preccurlyeq A^E x \preccurlyeq b^E + \delta\,\beta^E, A^I x \preccurlyeq b^I + \delta\,\beta^I\}. \tag{2.2}$$

---

2.1. Removal of implicit 0s is an idealization when dealing with a numerical solution algorithm. But at least in principle, the implicit 0s in (2.1) can be determined by solving a linear program for each of the $x_i$, of the form $\max x_i$ subject to $x \succcurlyeq 0$ and $A\,x \preccurlyeq b$.

The tolerances are only on the values of the constraints, not on their structure. In all our results below, the allowable error vectors $\beta^E$, $\beta^I$ can be any positive vectors, but to get interesting results, one should take them equal to $|b^E|$, $|b^I|$, with the exception of any elements of $b^E$, $b^I$ that are 0, in which case the corresponding elements of $\beta^E$, $\beta^I$ are set to an appropriate positive constant, dictated by the application. By writing the equality constraints as $|A^E x - b^E| \preccurlyeq \delta \, \beta^E$, it is clear that they are stronger than (imply) $\|A^E x - b^E\|_1 \leqslant \delta \, \|\beta^E\|_1$. Similarly, the inequality constraints imply $\|A^I x - b^I\|_1 \leqslant \delta \, \|\beta^I\|_1$.

For simplicity we have used a single tolerance $\delta$ in (2.2), whereas in reality a number of different $\delta$ might be needed, e.g. for equalities vs. inequalities, 0 vs. non-0 elements of $b$, etc. The extension to an arbitrary number of different $\delta$ is straightforward.

Having a tolerance on equalities is *necessary* for the concentration results to hold for all $n$ beyond a certain point. Conceptually, tolerances on inequalities are not really necessary. If they are omitted, the only thing that changes in the results of §3, which address the general case, is the value of the constant $\vartheta_\infty$ below. For the results of §4, where, essentially, we only have equality constraints, tolerances on inequalities are more a matter of allowing simple analytical results (see Lemma 4.10 and Example 4.7 in §4.3). Quite apart from the above, as already pointed out, tolerances may also be regarded as reflecting some uncertainty in the values $b$ of the constraints.

Clearly, vectors sufficiently close to $\varphi^*$ should belong to $\mathcal{C}(\delta)$:

**Proposition 2.1.** *Given the constraints ( 2.2), let $\beta_{\min}$ denote the smallest element of $\beta$ and define*

$$\vartheta_\infty = \max \, (\beta^E_{\min} / \|\|A^E\|\|_\infty, \beta^I_{\min} / \|\|A^I\|\|_\infty),$$

*where the matrix norm $\|\| \cdot \|\|_\infty$ is the maximum of the $\ell_1$ norms of the rows. Then for any $\delta > 0$, any $x \in \mathbb{R}^m$ s.t. $\|x - \varphi^*\|_\infty \leqslant \delta \, \vartheta_\infty$ belongs to $\mathcal{C}(\delta)$.*

We can assume that $\vartheta_\infty \leqslant 1$, otherwise some constraints are vacuous and can be eliminated.

**Example 2.2.** Suppose a die is tossed $n$ times and we are told just that the mean result was 4.5 (the "Brandeis dice" problem, [Jay83]). The matrices and vectors expressing this information or constraints are $A^E = [1, 2, 3, 4, 5, 6]$, $b^E = [4.5]$, and the MaxEnt p.d. given this information is $\varphi^* = (0.05435, 0.07877, 0.11416, 0.16545, 0.23978, 0.\,34749)$. Here we have $\beta^E = |b^E| = 4.5 = \beta^E_{\min}$, and the $\ell_1$ norm of the only row of $A^E$ is 21, hence $\|\|A^E\|\|_\infty = 21$. Thus $\vartheta_\infty = 0.2143$ in this case. If we take $\delta = 0.01$, Proposition 2.1 says that any $x \in \mathbb{R}^6$ s.t. $\max_i |x_i - \varphi^*_i| \leqslant 0.002143$ belongs to the set

$$\mathcal{C}(0.01) \, = \, \{x \in \mathbb{R}^6 \, | \, 4.455 \leqslant x_1 + 2\,x_2 + 3\,x_3 + 4\,x_4 + 5\,x_5 + 6\,x_6 \leqslant 4.545\}.$$

## 2.3 Tolerances and the MaxEnt p.d.

Another aspect of introducing tolerances is that $H^*$ is the maximum of the entropy over the domain $\mathcal{C}(0)$, with no tolerances on the constraints. However, as $\delta$ increases from 0, the maximum of $H$ over $\mathcal{C}(\delta)$ will generally shift from $\varphi^* = \varphi^*(0)$ to another point $\varphi^*(\delta)$, at which the entropy may attain a value $H^*(\delta)$ even greater than $H^* = H^*(0)$. So if $\delta$ is allowed to be too large with respect to $\vartheta$, we cannot expect concentration to still occur in a region of "size $\vartheta$" around $\varphi^*(0)$.

Below we show that the difference of the entropies of all frequency vectors in $F_n \cap \mathcal{C}(\delta)$ from $H^*(0)$ can be bounded by a *linear function* of $\delta$, so that by choosing a small enough $\delta$ we can ensure concentration in a $\vartheta$-sized region around $\varphi^*(0)$.

First we present some basic properties of the MaxEnt p.d. $\varphi^* = \varphi^*(0)$. Since all elements of $\varphi^*$ are positive (recall assumption 1 in §2.1), they can be written as

$$\varphi_j^* = e^{-(\lambda_0 + \lambda \cdot A_{\cdot j})} = \frac{1}{Z(\lambda)} e^{-\lambda \cdot A_{\cdot j}}, \quad \text{where} \quad Z(\lambda) = e^{\lambda_0} = \sum_k e^{-\lambda \cdot A_{\cdot k}} \tag{2.3}$$

(see e.g. [Jay03], [BV04]), where $\lambda_0$ is the Lagrange multiplier corresponding to the normalization constraint $\sum_i x_i = 1$ in (2.1), and $\lambda = (\lambda_1, ..., \lambda_{\ell_E + \ell_I})$ are the multipliers corresponding to the rest of the constraints. The positivity of $\varphi^*$ ensures that the multipliers exist and are finite. In (2.3) $\lambda \cdot A_{\cdot j}$ is the dot product of $\lambda$ and the $j$th column $A_{\cdot j}$ of $A = [A^E, A^I]^T$, the block column matrix formed by $A^E$ and $A^I$. The normalization factor $Z(\lambda)$ is sometimes referred to as the "partition function"[2.2].

$\varphi^*$ satisfies some of the inequality constraints with equality, and these are known as *binding* (or *active*) at the point $\varphi^*$, and some with strict inequality. Once $\varphi^*$ is known we can distinguish these two types of constraints and refine our notation to $A^I = [A^{\mathrm{BI}}, A^{\mathrm{NI}}]^T$. Then we can write

$$[A^{\mathrm{BI}}, A^{\mathrm{NI}}]^T \varphi^* + s^* = (b^{\mathrm{BI}}, b^{\mathrm{NI}})^T, \qquad s^* \succcurlyeq 0, \tag{2.4}$$

where the *slack* vector $s^*$ has 0s in the elements corresponding to $b^{\mathrm{BI}}$ and its other elements are positive.

Returning to the multipliers, it is known that if $\lambda_i$ corresponds to an inequality constraint that is non-binding at $\varphi^*$, then $\lambda_i = 0$ (e.g. [BV04], §5.5), and in that case this constraint plays no role in the solution, that is the $i$th row of $A$ does not appear in (2.3). So if we partition $\lambda$ as $(\lambda^E, \lambda^{\mathrm{BI}}, \lambda^{\mathrm{NI}})$, (2.3) can be written as

$$\varphi_j^* = e^{-(\lambda_0 + \lambda^E \cdot A_{\cdot j}^E + \lambda^{\mathrm{BI}} \cdot A_{\cdot j}^{\mathrm{BI}})}, \quad \text{where} \quad \lambda^{\mathrm{BI}} \succcurlyeq 0. \tag{2.5}$$

Here $\lambda^{\mathrm{BI}} \succcurlyeq 0$ because $\partial H^* / \partial b_j^{\mathrm{BI}} = \lambda_j^{\mathrm{BI}}$, and this must be $\geqslant 0$ since $H^*$ can't decrease if we increase $b_j^{\mathrm{BI}}$. The elements of $\lambda^E$ have an analogous interpretation but can be positive or negative.

In the rest of the paper *we treat the values of the multipliers $\lambda_0, \lambda^E, \lambda^{\mathrm{BI}}$ as known*. Since $\varphi^*$ is known, $(\lambda_0, \lambda^E, \lambda^{\mathrm{BI}})$ can be determined by taking logs of both sides of (2.5) and solving a linear system (which has at least as many equations as unknowns).

The values of the Lagrange multipliers can be used to bound the value of $H(\cdot)$ over all frequency vectors in $\mathcal{C}(\delta)$ in terms of $H^*(0)$ and $\varphi^*(0)$:

**Proposition 2.3.** *Let $(\lambda^E, \lambda^{\mathrm{BI}})$ be the vector of Lagrange multipliers in (2.5), corresponding to the solution $H^*(0), \varphi^*(0)$ of the entropy maximization problem (2.1). Set*

$$\Lambda^* \triangleq |\lambda^E| \cdot \beta^E + \lambda^{\mathrm{BI}} \cdot \beta^{\mathrm{BI}} > 0.$$

*Then for any $\delta > 0$, for all $f \in F_n \cap \mathcal{C}(\delta)$*

$$H(f) \leqslant H^*(0) + \Lambda^* \delta - D(f \| \varphi^*(0)).$$

The notation $\Lambda^*$ is a reminder that this quantity is computable after solving a *single* entropy maximization problem plus a linear system.

By considering the limit as $n \to \infty$, we can conclude from Proposition 2.3 that if $\varphi^*(\delta)$ is the density vector maximizing the entropy over $\mathcal{C}(\delta)$ and $H^*(\delta)$ is the corresponding maximum value, then $H^*(\delta) \leqslant H^*(0) + \Lambda^* \delta$. [This last result can also be obtained under more general conditions by "sensitivity analysis" methods in convex optimization, see e.g. §VII.3.3 of [HUL96]. The advantage of our more specialized derivation is the inclusion of the $D(\cdot \| \cdot)$ term.]

---

[2.2]. Here we are using Lagrange multipliers formally, not advocating them for actually computing $\varphi^*$.

We will use Proposition 2.3 in the combinatorial proofs of concentration results in §3 and in the probabilistic proofs of §4 as well.

# 3 Concentration, general case

The main results of this section are (a) the determination of the fraction of sequences satisfying a constraint up to stated tolerances (Theorem 3.14) and (b), the subsequent computation of the number $N(\delta, \varepsilon, \vartheta)$ appearing in the concentration statement $EC$ of §1, under arbitrary linear constraints as described in §2.1 (Theorem 3.15). The derivation is based on tight Stirling-type approximations to the multinomial coefficient, tight relationships between entropy and $\ell_1$ norm, and estimates for the number of lattice points in convex bodies. (In §4 we will calculate the number $N(\delta, \varepsilon, \vartheta)$ for the special case of equality constraints in a very different manner.)

We begin by introducing the *frequency* vector $f^*$ corresponding to the MAXENT probability vector $\varphi^*$, necessary when we want a discrete optimal solution to the maximum entropy problem, and then present a simple 'reference' or 'benchmark' concentration result against which we measure the improved results that follow. To simplify the notation, here we denote the $\varphi^*(0)$ and $H^*(0)$ of §2.3 simply by $\varphi^*, H^*$.

## 3.1 The MAXENT frequency vector

Let $F_n$ be the set of all frequency vectors $f = (\nu_1/n, ..., \nu_m/n)$, where the counts $\nu_i$ sum to $n$. Define the rounding of $x \in \mathbb{R}_+$ to an integer $[x]$ in the usual way, so that it satisfies $|x - [x]| \leqslant 1/2$. Given $n \in \mathbb{N}$, from the MAXENT probability vector $\varphi^*$ we derive a count vector $\nu^*$ and a frequency vector $f^*$ by a process of *rounding* and *adjusting* (recall that all elements of $\varphi^*$ are positive):

**Definition 3.1.** *Given $\varphi^*$ and $n \geqslant m$, let $\tilde{\nu} = [n\,\varphi^*]$ and $d = \sum_i \tilde{\nu}_i - n$. If $d = 0$, set $\nu^* = \tilde{\nu}$. Otherwise, if $d < 0$, add 1 to $|d|$ elements of $\tilde{\nu}$ that were rounded down, and if $d > 0$, subtract 1 from $|d|$ elements that were rounded up. Let the resulting vector be $\nu^*$, and define $f^* = \nu^*/n$, $f^* \in F_n$.*

Unlike $\varphi^*$, the vectors $\nu^*$ and $f^*$ depend on $n$, but we will not indicate this explicitly to avoid burdensome notation. The adjustment of $\tilde{\nu}$ in Definition 3.1 when $d \neq 0$ ensures that the result $\nu^*$ sums to $n$, so $f^*$ is a proper frequency vector. This adjustment is always possible: if $d < 0$ there must be at least $|d|$ elements of $n\,\varphi^*$ that were rounded to their floors, and if $d > 0$ to their ceilings, and $d \leqslant \lfloor m/2 \rfloor$.

**Example 3.2.** Continuing Example 2.2, as already mentioned the MAXENT p.d. is

$$\varphi^* = (0.05435, 0.07877, 0.11416, 0.16545, 0.23978, 0.\,34749).$$

If $n = 1000$, $[1000\,\varphi^*] = [(54.35, 78.77, 114.16, 165.44, 239.77, 347.49)]$. Four elements are rounded down and two up, resulting in $(54, 79, 114, 165, 240, 347)$. This sums to 999, so $d = -1$. Thus we can take $f^* = (55, 79, 114, 165, 240, 347)/1000$. If $n = 1003$, $[1003\,\varphi^*] = ([54.52, 79.01, 114.502, 165.94, 240.49, 348.54])$, and with four elements rounded up, $d = 1$. $f^*$ is not unique: we can take it to be $(55, 79, 115, 166, 240, 348)/1003$, $(55, 79, 115, 165, 240, 349)/1003$, etc. In the rest of the paper, all results involving the MAXENT frequency vector $f^*$ hold for any $f^*$ constructed according to Definition 3.1.

The importance of $\nu^*$ and $f^*$ is that when dealing with a discrete problem, we want to be able to *actually exhibit* a discrete solution having the desired properties, e.g. Theorem 3.4 or Theorem 3.17. Having just the MAXENT *probability* vector $\varphi^*$ will not do. The MAXENT *frequency* vector $f^*$ is a reasonable approximation (not unique) to the vector in $F_n$ that maximizes $H(f)$.

By construction, $f^*$ is close to $\varphi^*$ both in norm and in entropy:

**Proposition 3.3.** *With $f^*$ constructed as in Definition 3.1,*

$$\|f^* - \varphi^*\|_\infty \leqslant \frac{1}{n}, \quad and \quad \|f^* - \varphi^*\|_1 \leqslant \frac{3\,m}{4\,n}.$$

*Further,*

$$H(f^*) \geqslant H^* - \frac{3\,m}{8\,n}\ln\frac{8\,n}{3} + \left(1 - \frac{3\,m}{8\,n}\right)\ln\left(1 - \frac{3\,m}{8\,n}\right).$$

## 3.2 A reference result

We first establish a result which (a) introduces some basic ideas and (b) with which we can compare the more sophisticated results that follow, so as to be able to assess the improvements. The proof of this 'reference' or 'benchmark' result borrows ideas from that of the conditional limit theorem, [CT06] Theorem 11.6.2. However here we use plain instead of relative entropies, we assess the difference between two vectors via $\ell_1$ norm instead of entropy, we take tolerances in the constraints into account, and we do not use probabilities at all, but numbers of realizations.

It is customary in the literature to consider frequency vectors that satisfy the constraints in terms of the *difference of their entropy* from the maximum entropy value $H^*$. Thus given an $\eta > 0$, one partitions $F_n \cap \mathcal{C}(\delta)$ into the sets

$$\begin{aligned}
\mathcal{A}_n(\delta, \eta) &= \{f \in F_n \cap \mathcal{C}(\delta), H(f) \geqslant (1-\eta)\,H^*\}, \\
\mathcal{B}_n(\delta, \eta) &= \{f \in F_n \cap \mathcal{C}(\delta), H(f) < (1-\eta)\,H^*\}.
\end{aligned}$$

As mentioned in the Introduction, a better measure for our purposes is the *norm of the deviation* of a frequency vector from the maximum entropy *vector $\varphi^*$*. Besides being more intuitive, this measure is also more stringent than difference in entropy: with the $\ell_1$ norm, if $\|f - f'\|_1 = 0$ it must be that $f = f'$; but $H(f) = H(f')$ does not imply $f = f'$. Also, a small difference in entropy does not imply a small norm of the difference; e.g. if $f'$ is a permutation of $f$, the entropies are the same but $\|f - f'\|_1$ can achieve its maximum value of 2. On the other hand, if the norm of the difference is small so is the difference of the entropies (details in §3.3).

Adopting the norm measure, given a $\vartheta > 0$ we will define the sets

$$\begin{aligned}
\mathcal{A}_n(\delta, \vartheta) &\triangleq \{f \in F_n \cap \mathcal{C}(\delta), \|f - \varphi^*\|_1 \leqslant \vartheta\}, \\
\mathcal{B}_n(\delta, \vartheta) &\triangleq \{f \in F_n \cap \mathcal{C}(\delta), \|f - \varphi^*\|_1 > \vartheta\}.
\end{aligned} \tag{3.1}$$

Define the number of realizations $\#f$ of a frequency vector $f$ to be the same as the number of realizations of the corresponding count vector $\nu$. Further, let the number of realizations of a *set* of frequency or count vectors be the sum of the numbers of realizations of its elements. We want to show that given any $\varepsilon > 0$, there is a number $N = N(\delta, \varepsilon, \vartheta)$ s.t. if $n \geqslant N$, all but a fraction $\varepsilon$ of the realizations/assignments that satisfy the constraints $\mathcal{C}(\delta)$ have frequency vectors in the set $\mathcal{A}_n(\delta, \vartheta)$:

$$\frac{\#\mathcal{A}_n(\delta, \vartheta)}{\#\mathcal{A}_n(\delta, \vartheta) + \#\mathcal{B}_n(\delta, \vartheta)} = \frac{\#\mathcal{A}_n(\delta, \vartheta)}{\#(F_n \cap \mathcal{C}(\delta))} \geqslant 1 - \varepsilon. \tag{3.2}$$

If we were speaking in terms of differences in entropy, to prove the analogue of (3.2) we would consider the set

$$\mathcal{A}_n(\delta, \eta/2) = \{f \in F_n \cap \mathcal{C}(\delta), H(f) \geqslant (1 - \eta/2)\,H^*\} \tag{3.3}$$

which is well-separated from $\mathcal{B}_n(\delta, \eta)$ in the sense that any vector in it has entropy at least $(\eta/2)\,H^*$ greater than that of any vector in $\mathcal{B}_n(\delta, \eta)$.

When we speak in terms of $\ell_1$ norm of the difference we will try to do something similar, starting with relations between entropy and $\ell_1$ norm that hold for all frequency vectors in $\mathcal{C}(\delta)$:

$$\forall f \in F_n \cap \mathcal{C}(\delta), \quad \begin{aligned} \|f - \varphi^*\|_1 \leqslant \zeta &\Rightarrow H(f) \geqslant H^* - \zeta \ln(m/\zeta), \\ \|f - \varphi^*\|_1 > \vartheta &\Rightarrow H(f) < H^* - \vartheta^2/2 + \Lambda^* \delta. \end{aligned} \tag{3.4}$$

The first implication is independent of $\delta$ and follows from the $\ell_1$ norm bound on entropy difference ([CT06], Theorem 17.3.3). For the second implication, we have $H(f) \leqslant H^*(0) + \Lambda^* \delta - D(f \| \varphi^*)$ from Proposition 2.3, and then $\|f - \varphi^*\|_1 > \vartheta$ and Pinsker's inequality imply $D(f \| \varphi^*) > \vartheta^2/2$.

Suppose we choose $\delta$ so that $\|f - \varphi^*\|_1 > \vartheta \Rightarrow H(f) < H^* - \vartheta^2/3$. By (3.4) this requires $\delta < \vartheta^2/(6 \Lambda^*)$. With such $\delta$, we have the sets

$$\begin{aligned} \mathcal{A}_n(\delta, \zeta) &= \{ f \in F_n \cap \mathcal{C}(\delta), \|f - \varphi^*\|_1 \leqslant \zeta \}, & H(f) \geqslant H^* - \zeta \ln(m/\zeta), \\ \mathcal{B}_n(\delta, \vartheta) &= \{ f \in F_n \cap \mathcal{C}(\delta), \|f - \varphi^*\|_1 > \vartheta \}, & H(f) < H^* - \vartheta^2/3. \end{aligned} \tag{3.5}$$

Now suppose we want $\zeta$ to be s.t. any $f$ in the first set has entropy at least $\vartheta^2/4$ greater than any $f$ in the second set. By (3.5), $\zeta$ must then satisfy $\zeta \ln(m/\zeta) \leqslant \vartheta^2/12$, which holds if $\zeta = \vartheta^3/12$, and, in addition, $m \leqslant (\vartheta^3/12) \, e^{1/\vartheta}$. So under these conditions the set $\mathcal{A}_n(\delta, \vartheta^3/12)$ is well-separated from the set $\mathcal{B}_n(\delta, \vartheta)$.

We also need to establish that $\mathcal{A}_n(\delta, \vartheta^3/12)$ is not empty, i.e. that there is an $f$ in $\mathcal{C}(\delta)$ with $\|f - \varphi^*\|_1 \leqslant \vartheta^3/12$. But Proposition 3.3 says that the $f^*$ of Definition 3.1 is such that $\|f^* - \varphi^*\|_1 \leqslant 3 \, m/(4 \, n)$, so $\|f^* - \varphi^*\|_1$ will not exceed $\vartheta^3/12$ if $n \geqslant 9 \, m/\vartheta^3$. By Proposition 2.1, $f^*$ will also be in $\mathcal{C}(\delta)$ if $n \geqslant 1/(\delta \, \vartheta_\infty)$. These two requirements on $n$ ensure that $\mathcal{A}_n(\delta, \vartheta^3/12)$ contains at least the vector $f^*$.

To complete the argument we use the very simple bounds on the number of realizations of a frequency vector given in [CT06], Theorem 11.1.3, based on the method of types: $e^{nH(f)}/(n+1)^m \leqslant \#f \leqslant e^{nH(f)}$, and the number of elements in $\mathcal{B}_n$ is no more than $(n+1)^m$ irrespective of $\delta, \vartheta$. Thus we obtain our first 'reference' or 'benchmark' theorem

**Theorem 3.4.** *With the sets $\mathcal{A}_n, \mathcal{B}_n$ defined in (3.1), for any $\vartheta > 0$ and $\delta < \vartheta^2/(6 \Lambda^*)$, where $\Lambda^*$ has been defined in Proposition 2.3, if*

$$n \geqslant \frac{9 \, m}{\vartheta^3} \quad and \quad m \leqslant \frac{\vartheta^3}{12} \, e^{1/\vartheta},$$

*then*

$$\frac{\#\mathcal{A}_n}{\#\mathcal{B}_n} \geqslant \frac{\#f^*}{\#\mathcal{B}_n} \geqslant \frac{e^{n\vartheta^2/4}}{(n+1)^{2m}}.$$

This is not a novel result, apart from the introduction of the tolerance $\delta$ and its interaction with $\vartheta$. It is to be compared with its refinement, Theorem 3.14, and its alternative version optimized for large $m$, Theorem 4.1.

The last inequality in Theorem 3.4 shows the price we pay for using the stricter $\|f - \varphi^*\|_1$ as the measure of closeness instead of the looser $|H(f) - H(\varphi^*)|$: if $\mathcal{A}_n$ had been defined as in (3.3), the exponent above would have been $n \, \eta \, H^*/2$; but by (3.4) if the norm measure is $> \vartheta$ and we ignore tolerances, the difference in entropies is only of the order of $\vartheta^2$. This seems to be the best that can be said, see Proposition 3.6 below. (Intuitively, $H$ is a differentiable function of $f$, so a second-order Taylor expansion around $\varphi^*$ shows that changes in $H$ near the maximum are quadratic in terms of the corresponding changes in $f$. See also the remarks after Lemma 4.10.)

Theorem 3.4 immediately implies the second benchmark result, which is to be compared with Theorems 3.15 and 4.4 presented later:

**Theorem 3.5.** *For all $\varepsilon, \vartheta > 0$ and $\delta < \vartheta^2 / (6\,\Lambda^*)$, where $\Lambda^*$ is as in Proposition 2.3, if*

$$n \geqslant \frac{1}{\delta\,\vartheta_\infty}, \quad n \geqslant \frac{9\,m}{\vartheta^3}, \quad n \geqslant \frac{8\,m\ln(n+1) + 4\ln(1/\varepsilon)}{\vartheta^2},$$

*and moreover $m \leqslant (\vartheta^3/12)\,e^{1/\vartheta}$, then*

$$\frac{\#\mathcal{A}_n}{\#\mathcal{B}_n} > \frac{1}{\varepsilon}.$$

An approximation to the solution of the third inequality on $n$ above is $n \approx (8\,m\ln(8\,m/\vartheta^2) + 4\ln(1/\varepsilon))/\vartheta^2$, so we may expect the second inequality to dominate unless $\varepsilon$ is very small.

The benchmark Theorems 3.4 and 3.5 can be improved in various ways; the improvements are the subject of this paper.

## 3.3  Entropy difference and $\ell_1$ norm

The results (3.4) say that if two frequency vectors are close in $\ell_1$ norm, their entropies are also close, but if they are far apart, so are their entropies. The following is an improvement of these results:

**Proposition 3.6.** *Given the vector $\varphi^*$, $\delta > 0$, $\zeta \leqslant \varphi^*_{\min}$, and $\vartheta > 0$, for any $f \in F_n \cap \mathcal{C}(\delta)$*

$$\|f - \varphi^*\|_1 \leqslant \zeta \;\Rightarrow\; H(f) \geqslant H^* - h_2(\varphi^*_{\max}, \varphi^*_{\min}, \zeta),$$
$$\|f - \varphi^*\|_1 > \vartheta \;\Rightarrow\; H(f) < H^* - c(\varphi^*)\,\vartheta^2 + \Lambda^*\,\delta.$$

*In the first bound, the function $h_2$ is*

$$h_2(x_1, x_2, \zeta) \triangleq (x_1 + \zeta/2)\ln(x_1 + \zeta/2) - x_1\ln x_1 + (x_2 - \zeta/2)\ln(x_2 - \zeta/2) - x_2\ln x_2.$$

*In the second bound $c(\varphi^*) \geqslant 1/2$ is the number*

$$c(\varphi^*) \triangleq \frac{1}{4\,(1 - 2\,\beta(\varphi^*))}\ln\frac{1 - \beta(\varphi^*)}{\beta(\varphi^*)},$$

*where $\beta(\varphi^*) \leqslant 1/2$ is a characteristic of $\varphi^*$:*

$$\beta(\varphi^*) \triangleq \max_{I \subset \{1, ..., m\}} \min\left(\textstyle\sum_{i \in I} \varphi^*_i, 1 - \sum_{i \in I} \varphi^*_i\right).$$

*If $\beta(\varphi^*) = 1/2$, $c(\varphi^*) \triangleq 1/2$.*

We note that the first bound holds for general probability vectors, i.e. $f$ does not have to be a frequency vector and $\varphi^*$ does not have to be the MaxEnt probability vector, but the second bound requires $f$ to be a frequency vector and $\varphi^*$ to be the MaxEnt vector. With respect to determining the number $\beta(p)$ given a probability vector $p = (p_1, ..., p_m)$, we remark that this is an NP-complete problem[3.1], but in practice it can be solved in pseudo-polynomial time using dynamic programming [GW98].

---

3.1. Given an instance of the classic Partition problem (problem SP12 in [GJ78]) with weights $w_1, ..., w_m \in \mathbb{N}$, form the probability vector with elements $w_i / w$, where $w = \sum_i w_i$. Then an algorithm that computes $\beta(p)$ also solves the Partition problem: there is a partition of the set $\{1, ..., m\}$ into two equal-weight subsets iff $\beta(p) = 1/2$.

The first bound of the Proposition relies on a result of [HY10], and is tight. The function $h_2$ is identically 0 when $\zeta = 0$, increases as $\zeta$ increases, and is convex for $\zeta \in [0, \varphi^*_{\min}]$[3.2]. The second bound uses the improvement of Pinsker's inequality in [OW05], which is also tight, but we do not claim that the second bound itself is tight. In this bound, the farther away $\varphi^*$ is from having a partition, the smaller is $\beta(\varphi^*)$ and the bigger than $1/2$ is $c(\varphi^*)$. Then the bound is a significant improvement over the 2nd bound in (3.4) that follows from Pinsker's inequality.

**Example 3.7.** Continuing Example 3.2, suppose we were also told that the frequencies were such that $f_1 + f_2 \leqslant 0.1$. Under these two constraints we find $\varphi^* = (0.0434, 0.0566, 0.1445, 0.1885, 0.246, 0.321)$, with $H^* = 1.6026$. Let the tolerance $\delta$ be 0. If we take $\zeta = \vartheta = 0.02$, the lower and upper bounds of (3.4) on $H(f)$ are 1.4608, 1.6024 respectively. The lower bound of Proposition 3.6 is 1.5922, and the upper bound is 1.6022 with $\beta(\varphi^*) = 0.0434$, $c(\varphi^*) = 0.84678$.

We will use Proposition 3.6 in §3.5 to bound the entropies of the frequency vectors in $\mathcal{A}_n$ from below and the entropies of those in $\mathcal{B}_n$ from above.

## 3.4  Number of lattice points

We will need to be able to claim that the set of frequency vectors that satisfy $\|f - \varphi^*\|_1 \leqslant \vartheta$ contains at least so many elements, and the following results on lattice (integral) points in convex sets will be useful:

**Proposition 3.8.** *Let $\mathbb{S}_m(r)$ be a sphere in $\mathbb{R}^m$, $m \geqslant 3$, of radius $r > \sqrt{m}/2$. Then the number $L(\mathbb{S}_m(r))$ of lattice points, i.e. points in $\mathbb{Z}^m$, inside and on this sphere satisfies*

$$\frac{\pi^{m/2}}{\Gamma(m/2+1)} r^m \left(1 - \frac{\sqrt{m}}{2\,r}\right)^m \leqslant L(\mathbb{S}_m(r)) \leqslant \frac{\pi^{m/2}}{\Gamma(m/2+1)} r^m \left(1 + \frac{\sqrt{m}}{2\,r}\right)^m.$$

This is a consequence of the following relation between lattice points and volume:

**Theorem 3.9. (J. M. Wills, [IKKN04], §3.2)** *Let $\mathbb{K}$ be an $m$-dimensional strictly convex body, $m \geqslant 3$, and let $r > \sqrt{m}/2$ be the radius of the largest sphere contained in it. If $L(\mathbb{K})$ is the number of lattice points in and on $\mathbb{K}$, then*

$$\mathrm{vol}(\mathbb{K}) \left(1 - \frac{\sqrt{m}}{2\,r}\right)^m \leqslant L(\mathbb{K}) \leqslant \mathrm{vol}(\mathbb{K}) \left(1 + \frac{\sqrt{m}}{2\,r}\right)^m.$$

The applicability of the theorem is limited by the requirement of strict convexity (e.g. a polytope is not strictly convex); still, it suffices to establish Proposition 3.8 by taking the body $\mathbb{K}$ to be the sphere $\mathbb{S}_m(r)$.

Now let $\varphi^*_{\min}$ denote the smallest element of $\varphi^*$. Using Proposition 3.8 we can establish:

**Proposition 3.10.** *If $m \geqslant 3$, $n \geqslant m/(2\,\vartheta)$, and $\vartheta \leqslant \sqrt{m}\,\varphi^*_{\min}$, the set of frequency vectors $\{f \in F_n, \|f - \varphi^*\|_1 \leqslant \vartheta\}$ contains at least*

$$L(n, m, \vartheta) = \frac{1}{\Gamma((m+1)/2)} \left(\frac{\pi}{m}\right)^{\frac{m-1}{2}} \left(\vartheta\,n - \frac{m}{2}\right)^{m-1}$$

_______________

[3.2]. The last two properties follow from the fact that the first two derivatives of $h_2$ w.r.t. $\zeta$ are always positive.

*elements.*

If $n > (m+2)/(2\,\vartheta)$, then $L(n, m, \vartheta)$ is an increasing function of $n$. We will use Proposition 3.10 in deriving Lemma 3.12 below.

## 3.5   Bounds on $\#\mathcal{A}_n$ and $\#\mathcal{B}_n$

Define the sets $\mathcal{A}_n$, $\mathcal{B}_n$ of frequency vectors in terms of a tolerance $\vartheta$ on the deviation from the maximum entropy vector $\varphi^*$ as in (3.1), repeated here for convenience:

$$\begin{aligned}
\mathcal{A}_n(\delta, \vartheta) &\triangleq \{f \in F_n \cap \mathcal{C}(\delta), \|f - \varphi^*\|_1 \leqslant \vartheta\}, \\
\mathcal{B}_n(\delta, \vartheta) &\triangleq \{f \in F_n \cap \mathcal{C}(\delta), \|f - \varphi^*\|_1 > \vartheta\}.
\end{aligned}$$

There is a basic relationship between the number of realizations of a frequency vector and its entropy, already given in its simplest form just before Theorem 3.4. Given $f \in F_n$, assume w.l.o.g. that $f_1, \ldots, f_\mu, \mu \geqslant 1$, are its non-zero elements. Then a tighter relationship between $\#f$ and $H(f)$ is

$$e^{-\frac{\mu}{12}} e^{nH(f)} \leqslant \frac{\#f}{S(f, \mu)} \leqslant e^{nH(f)}, \quad \text{where} \quad S(f, \mu) \triangleq \frac{1}{(2\,\pi\,n)^{\frac{\mu-1}{2}}} \frac{1}{\sqrt{f_1 \cdots f_\mu}}. \tag{3.6}$$

(3.6) follows from the bounds on the factorial in [Fel68], §II.9, eq. (9.15), related to Stirling's approximation; see also [CK11], problem 2.2. Comparing (3.6) with the simple bounds given above Theorem 3.4 we see that the left inequality is improved by a factor of order $n^{\mu/2+1}$ and the right by a factor of order $n^{\mu/2-1}$.

We now establish an upper bound on the number of realizations of $\mathcal{B}_n$ and a lower bound on the number of realizations of $\mathcal{A}_n$. Both bounds have the form of a polynomial in $n$ times an exponential in $n$; the polynomial factors match, and therefore cancel when we take the ratio $\#\mathcal{A}_n / \#\mathcal{B}_n$. The bound on $\#\mathcal{A}_n$ involves a parameter $\alpha$, and what allows the polynomial factors to match is the condition $\alpha\,\vartheta < \varphi^*_{\min}$ which appears in Lemma 3.12 below.

**Lemma 3.11.** *Given the* MaxEnt *vector $\varphi^*$ and any $\delta, \vartheta > 0$,*

$$\#\mathcal{B}_n(\delta, \vartheta) \leqslant 4\sqrt{2\,\pi}\,(1/2 + 1/\sqrt{n})^m\, n^{\frac{m-1}{2}}\, e^{n(H^* + \Lambda^* \delta - c(\varphi^*)\vartheta^2)},$$

*where the number $c(\varphi^*)$ has been defined in Proposition 3.6.*

To state the lower bound on $\#\mathcal{A}_n$ we introduce a parameter $\alpha \in (0, 1)$, used to define a subset $\mathcal{A}_n(\delta, \alpha\,\vartheta)$ of $\mathcal{A}_n(\delta, \vartheta)$ which is well-separated from $\mathcal{B}_n(\delta, \vartheta)$; recall §3.2. We will end up optimizing over this parameter in the main result, Theorem 3.15.

**Lemma 3.12.** *Given the* MaxEnt *vector $\varphi^*$, $\delta, \vartheta > 0$, and some $\alpha \in (0, 1)$, if*

$$m \geqslant 3, \quad n \geqslant \frac{m}{2\,\alpha\,\vartheta}, \quad \text{and} \quad \alpha\,\vartheta < \min\Big(\varphi^*_{\min}, \frac{2}{\sum_i 1/\varphi^*_i}, \delta\,\vartheta_\infty\Big),$$

*where $\vartheta_\infty$ has been defined in Proposition 2.1, we have*

$$\begin{aligned}
\#\mathcal{A}_n(\delta, \vartheta) &\geqslant \frac{e^{-\frac{m}{12}}(2\,m)^{-\frac{m-1}{2}}}{\Gamma((m+1)/2)\,\sqrt{\varphi^*_1 \cdots \varphi^*_m}}\Big(1 - \frac{\alpha\,\vartheta}{2}\sum_i 1/\varphi^*_i\Big)\Big(\alpha\,\vartheta - \frac{m}{2\,n}\Big)^{m-1} \\
&\quad n^{\frac{m-1}{2}} e^{n(H^* - h_2(\varphi^*_{\max}, \varphi^*_{\min}, \alpha\vartheta))},
\end{aligned}$$

*where the function $h_2$ has been defined in Proposition 3.6.*

The detailed proof is in the Appendix, but a main ingredient of the bound on $\#\mathcal{A}_n(\delta, \vartheta)$ is the product of $\#f$, given by (3.6), and the number of lattice points in the set $\mathcal{A}_n(\delta, \alpha\,\vartheta)$, given by Proposition 3.10. This product gives rise to the factor $n^{(m-1)/2}$, which ends up cancelling the same factor in Lemma 3.11 when we take the ratio $\#\mathcal{A}/\#\mathcal{B}$ in Theorem 3.14 below.

The parameter $\alpha$ affects the lower bound on $n$ and has a complicated effect on $\#\mathcal{A}_n(\delta, \vartheta)$. We deal with this in the next section.

## 3.6  Concentration around, and at the MaxEnt vector

From the above two lemmas, the ratio $\#\mathcal{A}_n/\#\mathcal{B}_n$ is bounded below by a factor dependent polynomially on $m$ and $\alpha\,\vartheta$, times a factor exponential in $n$, of the form $e^{n\,\psi(\delta,\vartheta,\alpha)}$. When everything else is fixed, we want $\#\mathcal{A}_n/\#\mathcal{B}_n$ to increase without bound as $n$ increases, so we will want the function $\psi(\delta, \vartheta, \alpha)$ to be positive. The following proposition says that this will be so if $\delta$ and $\alpha$ are not too large:

**Proposition 3.13.** *Given $\delta, \vartheta > 0$, the function*

$$\psi(\delta, \vartheta, \alpha) \triangleq c(\varphi^*)\,\vartheta^2 - h_2(\varphi^*_{\max}, \varphi^*_{\min}, \alpha\,\vartheta) - \Lambda^*\,\delta,$$

*where $\Lambda^*$ has been defined in Proposition 2.3 and $c(\varphi^*)$ in Proposition 3.6, is positive if*

$$\delta < \frac{c(\varphi^*)}{\Lambda^*}\,\vartheta^2 \quad and \quad \alpha < \frac{\varphi^*_{\min}\big(c(\varphi^*)\vartheta^2 - \Lambda^*\,\delta\big)}{\vartheta\,h_2(\varphi^*_{\max}, \varphi^*_{\min}, \varphi^*_{\min})}.$$

*For fixed $\delta, \vartheta$, $\psi(\alpha) \nearrow$ as $\alpha \searrow$.*

Our first main result has to do with concentration *around* the MaxEnt probability *vector* $\varphi^*$, i.e. in the set $\mathcal{A}_n(\delta, \vartheta)$. It comes in two parts. The first is an analogue of the benchmark Theorem 3.4, and gives a lower bound on the ratio $\#\mathcal{A}_n/\#\mathcal{B}_n$:

**Theorem 3.14.** *Given $m \geqslant 3$, and $\delta, \varepsilon, \vartheta > 0$, define the constants*

$$\begin{aligned}
C_0(m) &\triangleq \frac{e^{-\frac{m}{12}}\,(m/2)^{-\frac{m-1}{2}}}{2\sqrt{2\,\pi}\,\Gamma((m+1)/2)\,\sqrt{\varphi^*_1\cdots\varphi^*_m}}, \\
C_1(m, \vartheta, \alpha) &\triangleq \frac{1 - (\alpha\,\vartheta/2)\sum_{1\leqslant i\leqslant m} 1/\varphi^*_i}{\big(1 + \sqrt{8\,\alpha\,\vartheta/m}\big)^m},
\end{aligned}$$

*where $\alpha \in (0, 1)$ is the parameter introduced in Lemma 3.12. Then*

$$\frac{\#\mathcal{A}_n(\delta, \vartheta)}{\#\mathcal{B}_n(\delta, \vartheta)} \geqslant C_0(m)\,C_1(m, \vartheta, \alpha)\,\Big(\alpha\,\vartheta - \frac{m}{2\,n}\Big)^{m-1} e^{n\psi(\delta,\vartheta,\alpha)}, \tag{3.7}$$

*where the function $\psi$ was defined in Proposition 3.13.*

The second main result follows from this and gives a bound on the least $n$ needed to get a particular degree of concentration. It should be compared to the benchmark Theorem 3.5. Because we are trying to minimize the required $n$, and, in particular, optimize over the parameter $\alpha$ which defines $\mathcal{A}_n(\delta, \alpha\,\vartheta)$, the result is not a bound on $n$ in a nice closed form, but more like an algorithm for computing the least $n$:

**Theorem 3.15.** *Continuing Theorem 3.14, define*

$$\lambda(m, \varepsilon, \vartheta, \alpha) \triangleq \ln 1 / \varepsilon - \ln C_0(m) \, C_1(m, \vartheta, \alpha) + (m-1) \ln 1 / (\alpha \, \vartheta).$$

*Let*

$$\delta < \frac{c(\varphi^*)}{\Lambda^*} \vartheta^2,$$

*where $c(\varphi^*)$ is as in Proposition 3.6 and $\Lambda^*$ as in Proposition 2.3, and set*

$$\hat{\alpha} \triangleq \min \left( \frac{\varphi^*_{\min}}{\vartheta}, \frac{2}{\vartheta \sum_i 1 / \varphi^*_i}, \frac{\delta \, \vartheta_\infty}{\vartheta}, \frac{\varphi^*_{\min} \big( c(\varphi^*) \vartheta^2 - \Lambda^* \delta \big)}{\vartheta \, h_2(\varphi^*_{\max}, \varphi^*_{\min}, \varphi^*_{\min})}, 1 \right).$$

*Finally, let*

$$N(\delta, \varepsilon, \vartheta) \triangleq \min_{\alpha \in (0, \hat{\alpha}]} \left( \max \left( \frac{m}{2 \, \alpha \, \vartheta}, \frac{\lambda(m, \varepsilon, \vartheta, \alpha)}{\psi(\delta, \vartheta, \alpha)} \right) + \frac{m}{\sqrt{2 \, \alpha \, \vartheta \, \psi(\delta, \vartheta, \alpha)}} \right).$$

*Then if $m \geqslant 3$ and $n \geqslant N(\delta, \varepsilon, \vartheta)$ we have for any $\varepsilon > 0$*

$$\frac{\# \mathcal{A}_n(\delta, \vartheta)}{\# \mathcal{B}_n(\delta, \vartheta)} \geqslant \frac{1}{\varepsilon}.$$

The result of this theorem is complicated. To make sense out of it, it is helpful to have some simple lower bounds on $N$. We find that

$$N(\delta, \varepsilon, \vartheta) \geqslant \max \left( \frac{m^3}{4}, \frac{m}{2 \, \vartheta_\infty \, \delta}, \frac{m \, h_2(\varphi^*_{\max}, \varphi^*_{\min}, \varphi^*_{\min})}{2 \, \varphi^*_{\min} \big( c(\varphi^*) \, \vartheta^2 - \Lambda^* \delta \big)} \right). \tag{3.8}$$

[These bounds are found by taking $m / (2 \, \alpha \, \vartheta)$ and replacing $\alpha$ with the 2nd, 3d, and 4th expressions in the definition of $\hat{\alpha}$, respectively. For the first bound note that $2 / \sum_i 1 / \varphi^*_i < 2 / m^2$, as the harmonic mean is < the arithmetic mean.]

The bounds of (3.8) on the least $n$ are to be compared with the conditions of Theorem 3.5. They will also prove useful in Example 4.15, where we compare the $N(\delta, \varepsilon, \vartheta)$ derived here to an $N(\delta, \varepsilon, \vartheta)$ derived using entirely different (probabilistic) techniques.

**Example 3.16.** In Example 3.7 we knew that the mean result was 4.5 and that $f_1 + f_2 \leqslant 0.1$. The matrices and vectors expressing this information or constraints are

$$A^E = [1, 2, 3, 4, 5, 6], \quad A^I = [1, 1, 0, 0, 0, 0], \quad b^E = [4.5], \quad b^I = [0.1].$$

From Proposition 2.1, $\vartheta_\infty = \max (4.5 / 21, 0.1 / 2) = 0.2143$. Further, $\varphi^* = (0.04339, 0.05661, 0.1445, 0.1885, 0.246, 0.321)$, and $\Lambda^* = 1.264$. Table 3.1 lists a few examples of the 'reference' $N(\delta, \varepsilon, \vartheta)$ established by Theorem 3.5 vs. the $N(\delta, \varepsilon, \vartheta)$ obtained from Theorem 3.15.

| $\delta$ | $\varepsilon$ | $\vartheta$ | $N_{\text{ref}}$ | $N$ |
|---|---|---|---|---|
| $5 \cdot 10^{-5}$ | $10^{-30}$ | 0.025 | $3.46 \cdot 10^6$ | $3.41 \cdot 10^5$ |
| $4 \cdot 10^{-5}$ | $10^{-6}$ | 0.02 | $6.75 \cdot 10^6$ | $4.37 \cdot 10^5$ |
| | $10^{-30}$ | | | $5.42 \cdot 10^5$ |
| $10^{-5}$ | $10^{-6}$ | 0.02 | | $1.56 \cdot 10^6$ |
| | $10^{-30}$ | | | $2.27 \cdot 10^6$ |
| | $10^{-6}$ | 0.01 | $5.4 \cdot 10^7$ | $1.75 \cdot 10^6$ |
| | $10^{-30}$ | | | $2.27 \cdot 10^6$ |
| $3 \cdot 10^{-6}$ | $10^{-30}$ | 0.005 | $4.32 \cdot 10^8$ | $9.68 \cdot 10^6$ |

**Table 3.1.** The reference $N$ of Theorem 3.5 and the $N$ of Theorem 3.15.

Theorem 3.15 is interesting, but in many situations it is not enough to know that an entire *set* of vectors around the MaxEnt probability vector $\varphi^*$ has most of the realizations; one would like to know something about a *specific* vector, $f^*$ for example. Our third main result concerns concentration *at* the MaxEnt frequency vector $f^*$ itself (recall Definition 3.1), and is much simpler than Theorem 3.15:

**Theorem 3.17.** *Given $\delta, \varepsilon, \vartheta > 0$, with $\delta < c(\varphi^*)\,\vartheta^2/\Lambda^*$, where $c(\varphi^*)$ has been defined in Proposition 3.6 and $\Lambda^*$ in Proposition 2.3, suppose that $n$ satisfies the conditions*

$$n \geqslant \frac{1}{\delta\,\vartheta_\infty}, \quad n \geqslant \frac{3\,m}{4\,\vartheta}, \quad n \geqslant \frac{(1.375\,m-1)\ln n + (1.2+1.5\,\vartheta)\,m + (1/2)\sum_i \ln f_i^* + \ln(4/\varepsilon)}{c(\varphi^*)\,\vartheta^2 - \Lambda^*\,\delta},$$

*where $\vartheta_\infty$ has been defined in Proposition 2.1. Then*

$$f^* \in \mathcal{C}(\delta) \quad and \quad \frac{\#f^*}{\#\mathcal{B}_n(\delta,\vartheta)} \geqslant \frac{1}{\varepsilon}.$$

This theorem says that for large enough $n$, the MaxEnt frequency vector $f^*$ *itself* has overwhelmingly more realizations than the entire set of vectors satisfying the constraints to the prescribed accuracy $\delta$ but differing from $\varphi^*$ by more than $\vartheta$ in norm.

Note that we cannot exclude *everything* around $f^*$: vectors close to it have comparable numbers of realizations, and if they are excluded from the set $\mathcal{A}$ but included in $\mathcal{B}$, concentration around $f^*$ will cease to hold. E.g. with $n$ even, $m = 2$, and no constraints, $\varphi^* = f^* = (1/2, 1/2)$ and $\#f^* = \binom{n}{n/2}$. But $\binom{n}{n/2}/\binom{n}{n/2\pm1} \to 1$ as $n$ increases. This is reflected in the theorem: the required $n \to \infty$ if $\vartheta \to 0$.

Comparing Theorem 3.17 with the reference Theorem 3.5, we see a marked improvement in the last two conditions on $n$ for the same choice of $\delta$. ($f^*$ depends on $n$, but the sum of the logs of the $f_i^*$ is always negative.)

## 3.7 Concentration around the uniform distribution

In the absence of any constraints in (2.1), besides non-negativity and normalization, the maximum entropy distribution is uniform. In this special case, the statement of Theorem 3.17 simplifies considerably, and we get the following corollary:

**Corollary 3.18.** *When the MaxEnt p.d $\varphi^*$ is uniform, if*

$$n \geqslant \frac{3\,m}{4\,\vartheta}, \quad n \geqslant \frac{(2.75\,m-2)\ln n - (\ln m - 3.1 - 3\,\vartheta)\,m + 2\ln(4/\varepsilon)}{\vartheta^2},$$

*then the approximately uniform MaxEnt frequency vector $u^*$ defined by*

$$u_i^* = \begin{cases} \lceil n/m \rceil / n, & 1 \leqslant i \leqslant n \bmod m, \\ \lfloor n/m \rfloor / n, & otherwise \end{cases}$$

*is such that*

$$\frac{\#u^*}{\#\{f \in F_n, \sum_i |f_i - 1/m| > \vartheta\}} \geqslant \frac{1}{\varepsilon}.$$

Laplace's famous "principle of indifference" or "principle of insufficient reason" says that in the absence of any knowledge that distinguishes among a number of possibilities we should adopt the uniform probability distribution (roughly speaking; the situation is actually quite a bit subtler, see [Jay03] Chapters 12 and 18). In our balls-and-bins paradigm, if we view each ball as a probability 'quantum' of size $1/n$, the object we construct by placing the balls in the bins is a discrete *probability distribution*, e.g. the $u^*$ of the Corollary[3.3].

The Corollary has easy-to-calculate implications for how compelling the Principle of Indifference is for given $m$ and $\varepsilon$. Of course, one could get even better bounds on the minimum $n$ needed by direct counting of all the possibilities, but this operation rapidly becomes less efficient with increasing $m$. For $m = 2$, the simplest case, we can directly calculate the minimum $n$; the following example gives a comparison.

**Example 3.19.** Suppose we consider all possible ways of constructing a 2-element discrete p.d. by placing $n$ probability quanta in two bins. The quanta are indistinguishable, but the bins are labelled '1' and '2'. With 5 quanta, a possible construction is $1, 2, 2, 2, 1$, meaning that first bin 1 receives a quantum, then bin 2, then bin 2 again, and again, and finally bin 1 gets another quantum. Now suppose we take $\vartheta = 0.1$. As we go through the $2^n$ constructions/assignments, if a construction results in the frequencies $u^* = (\lfloor n/2 \rfloor / n, \lceil n/2 \rceil / n)$ put it in the set $\mathcal{U}^*$, but if it results in frequencies such that either $f_1$ or $f_2$ differs from 0.5 by more than 0.05, put it in the set $\mathcal{B}$. Corollary 3.18 says that in the end the size of $\mathcal{U}^*$ will be larger than that of $\mathcal{B}$ by the factor shown in Table 3.2. In this simple two-bin case it is feasible to calculate the required $N$ exactly, by noting that the last expression in Corollary 3.18 translates to

$$\frac{\binom{N}{\lfloor N/2 \rfloor}}{2^N - \sum_{N(1-\vartheta)/2 \leqslant k \leqslant N(1+\vartheta)/2} \binom{N}{k}} \geqslant \frac{1}{\varepsilon}. \tag{3.9}$$

Table 3.2 shows the $N$ of the Corollary and the above exact $N$.

| $N$ | $|\mathcal{U}^*|/|\mathcal{B}|$ | $N_{\text{exact}}$ |
|---|---|---|
| 3694 | 1 | 410 |
| 4200 | 10 | 910 |
| 4699 | $10^2$ | 1350 |
| 5687 | $10^4$ | 2280 |
| 6664 | $10^6$ | 3200 |

**Table 3.2.** The $N$ of Corollary 3.18 and the ratio of the number of assignments to the two bins that result in $(f_1, f_2) = u^*$ to those that result in either $f_1$ or $f_2$ differing by more than 0.05 from 0.5. The last column is the $N$ obtained from (3.9).

So, the caveats mentioned under 'Relevance' at the end of §1 notwithstanding, here is some intuitive *and* quantitative support for the principle of indifference: with about 5000 balls/quanta, the corollary tells us that only about 1 in 1000 of all the possible constructions will result in a distribution significantly different from $(0.5, 0.5)$.

# 4 Concentration with no binding inequality constraints

Suppose that the constraints (2.1) consist only of equalities, or that all the inequalities turn out to be non-binding at the solution $\varphi^*$. Then concentration occurs earlier than implied by Theorem 3.14, because it is possible to show that the required $n$ no longer depends on $m$, but on the *number of constraints*, which is typically much smaller[4.1]. However, for the improvement to be noticeable in practice $m$ may have to be large. Perhaps surprisingly, it also becomes possible to have $m > n$, unlike in §3. We explore this in §4.4.

---

3.3. This is called the "Wallis derivation" of MaxEnt in [Jay03]. Jaynes' formulation also involves a team of monkeys, but here the authors thought they could manage well enough by themselves.

4.1. But not always. For example, to describe a regular hexagon in $\mathbb{R}^2$, one needs 6 inequalities, one per face.

There are two ingredients to this result, both probabilistic in nature, unlike the results of §3. So some of the simplicity and directness of the counting arguments of §3 is lost. The first ingredient is that under the MaxEnt p.d. $\varphi^*$ and *general* linear constraints, including any kind of inequalities, the ratio of the *probabilities* of $\mathcal{B}_n$ and $\mathcal{A}_n \cup \mathcal{B}_n$ is exponentially small in $n$. The second ingredient is that when the restriction to just equalities and non-binding inequalities is imposed, the ratio of probabilities under $\varphi^*$ translates to a ratio of *numbers of realizations*. Thus this ratio is shown to be exponentially small in $n$ as well.

In this section we will consider a generalization of the sets $\mathcal{A}_n, \mathcal{B}_n$ of (3.1) in §3.2 to

$$
\begin{aligned}
\mathcal{A}_n(\delta, \vartheta) &= \{f \in F_n \cap \mathcal{C}(\delta), \|f - \varphi^*\| \leqslant \vartheta\}, \\
\mathcal{B}_n(\delta, \vartheta) &= \{f \in F_n \cap \mathcal{C}(\delta), \|f - \varphi^*\| > \vartheta\},
\end{aligned}
\tag{4.1}
$$

where $\|\cdot\|$ denotes either the $\ell_1$ or the $\ell_2$ norm, and obtain results for both norms. The result for the Euclidean norm will turn out to be somewhat stronger (§4.4).

## 4.1  Concentration around the MaxEnt p.d.

The main results are Theorem 4.1 and 4.4 below. We outline how they follow from three basic results established later. First, Lemma 4.10 in §4.3 gives a lower bound for $\Pr_{\varphi^*}(F_n \cap \mathcal{C}(\delta))$. Next, Lemma 4.14 in §4.4 gives an upper bound on $\Pr_{\varphi^*}(\|f - \varphi^*\| > \vartheta)$, which applies a fortiori to $\Pr_{\varphi^*}(\mathcal{B}_n(\delta, \vartheta))$. Using these bounds, the ratio of the probability of the set $\mathcal{B}_n(\delta, \vartheta)$ to that of the set $\mathcal{A}_n(\delta, \vartheta) \cup \mathcal{B}_n(\delta, \vartheta) = F_n \cap \mathcal{C}(\delta)$ under the MaxEnt p.d. $\varphi^*$ is seen to be exponentially small in $n$:

$$
\frac{\Pr_{\varphi^*}(\mathcal{B}_n(\delta, \vartheta))}{\Pr_{\varphi^*}(\mathcal{A}_n(\delta, \vartheta) \cup \mathcal{B}_n(\delta, \vartheta))} \leqslant \frac{1}{\chi(\rho_1^2(\delta)\, n\,|\ell) - c\,\mu_3\,\ell^d/\sqrt{n}} \cdot
\begin{cases}
e^{-(\vartheta\sqrt{n} - s_1)^2/2}, & \ell_1 \text{ norm} \\
e^{-(\vartheta\sqrt{n} - \sqrt{1 - s_2})^2/2}, & \ell_2 \text{ norm}
\end{cases}.
\tag{4.2}
$$

Here $\chi(z|\ell)$ is the c.d.f. of a chi-squared distribution with $\ell$ degrees of freedom (see [AS72] §26.4), and the other quantities appearing in the bound are functions of $\varphi^*$, such as $s_1 = s_1(\varphi^*)$ and $s_2 = s_2(\varphi^*)$ defined in Lemma 4.14 and $\mu_3$ defined in Lemma 4.10, or of the constraints, such as $\rho_1(\delta)$ defined in Lemma 4.10, or absolute constants such as $c, d$, also defined in Lemma 4.10.

The bound (4.2) is valid under *general* linear constraints, including inequalities, so $\ell$ here is the number of all constraints as defined in §2.1.

The next step is to apply (4.7) of §4.2, to translate the ratio of probabilities (4.2) to a ratio of numbers of realizations. We then get the following theorem, which should be compared to Theorems 3.4 and 3.14:

**Theorem 4.1.** *For any $\delta, \vartheta > 0$ we have*

$$
\frac{\#\mathcal{B}_n(\delta, \vartheta)}{\#\mathcal{A}_n(\delta, \vartheta) + \#\mathcal{B}_n(\delta, \vartheta)} \leqslant \frac{e^{2\Lambda^*\delta n}}{\chi(\rho_1^2(\delta)\, n\,|\ell) - c\,\mu_3\,\ell^d/\sqrt{n}} \cdot
\begin{cases}
e^{-(\vartheta\sqrt{n} - s_1)^2/2}, & \ell_1 \text{ norm} \\
e^{-(\vartheta\sqrt{n} - \sqrt{1 - s_2})^2/2}, & \ell_2 \text{ norm}
\end{cases},
$$

*where $\Lambda^* = |\lambda^E| \cdot \beta^E > 0$ has the same definition as in Proposition 2.3 of §2.3, but here we have only equality constraints. (Recall from §2.3 that the vector $\lambda^E$ of Lagrange multipliers is assumed to be known.)*

Unlike (4.2), the bound of Theorem 4.1 is valid *only* when there are no binding inequalities; we explore the difficulties presented by inequality constraints that are binding in §4.2.

**Remark 4.2.** How does Theorem 4.1 compare to its analogues Theorems 3.4 and 3.14? Before getting into details, we give a simple informal argument showing that in general, when $m$ is large, the $N$ obtainable from Theorem 4.1 will be smaller than that of Theorems 3.4 and 3.14.

Let $N_0$ be the solution of $\chi(\rho_1^2(\delta)\, n\,|\ell) - c\,\mu_3\,\ell^d/\sqrt{n} = 0$; this number is independent of $m$. If $n$ is at least as large as some multiple of $N_0$, we can assume that the denominator of the r.h.s. in Theorem 4.1 is larger than some constant $c_1 \in (0,1)$. Then if $\delta$ and $\vartheta$ satisfy the condition $\delta < \vartheta^2/(4\,\Lambda^*)$, the numerator of that r.h.s. is of the form $c_2\,e^{-c_3 n}$, hence the entire r.h.s. is of the form $c_4\,e^{-c_3 n}$, where the constants are independent of $m$. In contrast, the $N$ of Theorem 3.14 cannot be any smaller than $m^3/4$ by the lower bound in (3.8), and for sufficiently large $m$ this will exceed the $n$ required to make $c_4\,e^{-c_3 n} \leqslant 1/\varepsilon$.

We said "in general" above because there is a hidden detail: $s_1$ and $\mu_3$ depend on the MaxEnt vector $\varphi^*$, which itself depends on the dimension $m$, so the $N_0, c_3, c_4$ above are not really entirely independent of $m$. On the other hand, despite its dependence on $\varphi^*$, $s_2$ is truly independent of $m$. So the $\ell_1$ result is more sensitive to $m$ than the $\ell_2$ result. Now Remark 4.11 below suggests that in some, rather pathological situations, $\mu_3$ can become large for large $m$, which would impose a lower bound on $n$ in terms of $m$. This would render both the $\ell_1$ and $\ell_2$ bounds of Theorem 4.1 less useful. Still, the second, final case in Example 4.15 shows that in 'ordinary' situations both $s_1$ and $\mu_3$ remain small even if $m$ is essentially infinite, and then the argument we gave above applies to both the $\ell_1$ and $\ell_2$ cases.

**Remark 4.3.** In practice we may want to consider a tolerance $\delta$ that depends on $n$. For example, if we obtain a sample of $n$ outcomes and observe a precise constraint, we may expect, if data are i.i.d. (or, much more generally, if the data obey the central limit theorem), that if we obtained a second independent sample, then the value of the constraint in the second sample can be expected to be different from its observed value in the first sample, but the distance will not be beyond order $1/\sqrt{n}$. Therefore, if we want to use the MaxEnt distribution as usual to make inferences about new, previously unseen data, i.e. about a second sample, it makes sense to impose the constraints only to a tolerance of order $1/\sqrt{n}$. For different reasons, it is also useful to let $\vartheta$ shrink with $n$: just as in standard concentration of measure where the radius of the smallest ball around the mean that gets nearly all of the probability shrinks with $n$, the same happens with entropy concentration. So it is interesting to determine the fastest rate at which we can let it shrink while still having concentration. We now analyze Theorem 4.1 in this light and compare it once again to the earlier results, Theorems 3.4 and 3.14. Set $\delta \asymp n^{-a}$ for some $a \geqslant 0$ ($a = 0$ yields constant $\delta$), and take $\vartheta \asymp n^{-b}$ for some $b \geqslant 0$[4.2]. From the definition (4.16) of the chi-squared distribution, it can be shown that for $\ell \geqslant 2$, if $w \leqslant 1$, then $\chi(w\,|\ell) \leqslant w$. Using this and the definition (4.17) and (4.12) of $\rho_1(\delta)$, it follows that $\chi(\rho_1^2(\delta)\, n\,|\ell) \asymp \min(1, \delta^2 n) \asymp \min(1, n^{1-2a})$. Therefore, to guarantee that the denominator in Theorem 4.1 does not become negative, we must set $a < 3/4$. If so, then this denominator is of order $\min(1, n^{1-2a})$. To make the numerator go to 0 with increasing $n$, we must have by the condition between $\vartheta$ and $\delta$ that $\vartheta$ is at least of order $\sqrt{\delta}$, i.e. $\vartheta \asymp n^{-a'/2}$ for some $0 \leqslant a' \leqslant a < 3/4$. With these choices the ratio in the theorem goes to 0 at rate $\max(1, n^{2a-1}) \exp(-C\,n^{1-a'})$, i.e.

$$\frac{\#\mathcal{B}_n(\delta,\vartheta)}{\#\mathcal{A}_n(\delta,\vartheta)} = O\Big(\frac{\#\mathcal{B}_n(\delta,\vartheta)}{\#\mathcal{A}_n(\delta,\vartheta) + \#\mathcal{B}_n(\delta,\vartheta)}\Big) = O(\exp(-C_0\,n^{1-a'} + C_1\,\lfloor 2\,a - 1\rfloor^+ \ln n)) \qquad (4.3)$$

for some constants $C_0, C_1$, where $\lfloor x\rfloor^+ = x$ if $x \geqslant 0$ and is 0 otherwise. In contrast, with $\vartheta$ again of order $n^{-a'/2}$, the asymptotic form of the bound given in benchmark Theorem 3.4 (with the fraction inverted to facilitate comparison), is immediately seen to be

$$\frac{\#\mathcal{B}_n(\delta,\vartheta)}{\#\mathcal{A}_n(\delta,\vartheta)} = O(\exp(-C_0'\,n^{1-a'} + C_1'\,m \ln n)), \qquad (4.4)$$

---

4.2. Here $\asymp$ has its usual meaning, $x_n \asymp y_n$ iff there exist constants $c_1, c_2 > 0$ such that for all $n$, $c_1\,x_n \leqslant y_n \leqslant c_2\,x_n$.

for some constants $C'_0, C'_1$, which cannot be competitive with (4.3) if $m \geqslant n^{1-a'}$ (assuming again that $s_1$ and $\mu_3$ do not depend on $m$). Theorem 3.14 does not readily allow an asymptotic analysis like (4.3) or (4.4) because of the complicated 'constants' $C_0$ and $C_1$ in (3.7). Yet we can see that for the result there to be meaningful at all, with $\vartheta$ of order $n^{-a'/2}$, we must have that $m = O(n^{1-a'/2})$, otherwise the third factor in (3.7) becomes negative. Hence both earlier results have constraints on $m$ in terms of $n$, *even if* the tolerances shrink to 0.

We now present our analogue of Theorems 3.5 and 3.15. This new theorem follows from Theorem 4.1 above; its statement involves a number of quantities that are functions of $\varphi^*$ or of the constraints, or absolute constants, as we have already described, and whose exact definitions are given later. Because we are attempting to minimize the required $N$, the result rests on the numerical solution of an inelegant equation involving the parameters $\delta, \varepsilon, \vartheta$ and quantities depending on $\varphi^*$, and, like Theorem 3.15, reads more like an algorithm for computing $N$:

**Theorem 4.4.** *Suppose that the constraints are such that no inequalities are binding at $\varphi^*$. Given $\delta, \varepsilon, \vartheta > 0$, assume that*

$$\delta < \vartheta^2 / (4\,\Lambda^*),$$

*where $\Lambda^*$ is as in Theorem 4.1.*

*Let $N_1 = N_1(\delta, \vartheta, \varepsilon)$ be the solution of the equation*

$$(0.5\,\vartheta^2 - 2\,\Lambda^*\delta)\,n - \vartheta\,s_1\,\sqrt{n} + \ln(\chi(\rho_1^2(\delta)\,n\,|\ell) - c\,\mu_3\,\ell^d/\sqrt{n}) = \ln(1/\varepsilon) - 0.5\,s_1^2,$$

*where $\chi(\cdot|\cdot)$ is the chi-squared distribution, $s_1 = s_1(\varphi^*)$ is defined in Lemma 4.14, $\ell$ is the number of constraints, and $\rho_1(\delta)$, $\mu_3$, and $c, d$ are defined in Lemma 4.10.*

*Further, let $N_2 = N_2(\delta, \vartheta, \varepsilon)$ be the solution of the similar equation*

$$(0.5\,\vartheta^2 - 2\,\Lambda^*\delta)\,n - \vartheta\,\sqrt{1 - s_2}\,\sqrt{n} + \ln(\chi(\rho_1^2(\delta)\,n\,|\ell) - c\,\mu_3\,\ell^d/\sqrt{n}) = \ln(1/\varepsilon) - 0.5\,(1 - s_2),$$

*where $s_2 = s_2(\varphi^*)$ is defined in Lemma 4.14.*

*Then if $n \geqslant N_1(\delta, \vartheta, \varepsilon)$ when $\mathcal{A}_n, \mathcal{B}_n$ are defined in terms of the $\ell_1$ norm in (4.1), or if $n \geqslant N_2(\delta, \vartheta, \varepsilon)$ when they are defined in terms of the $\ell_2$ norm, we have*

$$\frac{\#\mathcal{B}_n(\delta, \vartheta)}{\#\mathcal{A}_n(\delta, \vartheta) + \#\mathcal{B}_n(\delta, \vartheta)} \leqslant \varepsilon.$$

*The solution $N_0$ to $\chi(\rho_1^2(\delta)\,n\,|\ell) - c\,\mu_3\,\ell^d/\sqrt{n} = 0$ is a lower bound on both $N_1$ and $N_2$.*

The bound on $\delta$ in terms of $\vartheta$ ensures that the coefficient of $n$ in the equations for $N_1$, $N_2$ is positive. As can be seen by referring back to Theorem 4.1, this bound is an expression of the dependence between $\delta$ and $\vartheta$ that we saw in §3.2, but with some differences. First, here the upper bound on the probability of the set $\mathcal{B}$ (Lemma 4.14 in §4.4) does not depend on $\delta$, but just on $\vartheta$, and is thus *unaffected* by the MAXENT p.d. shifting from $\varphi^*(0)$ to $\varphi^*(\delta)$. Nevertheless, for a given $\vartheta$, increasing $\delta$ allows exponentially large fluctuations in the ratio of probabilities to realizations in (4.7), and this counteracts the fact that the probability of $\mathcal{B}$ is exponentially small in $\vartheta$. However, the origin of these fluctuations is the 'sensitivity' result of Proposition 2.3; and this same result underlies the bound on $\delta$ in §3.

Example 4.15 in §4.4 illustrates the theorem.

**Remark 4.5.** Just as in Remark 4.3, when exploring Theorem 4.4 it is again useful to consider a tolerance $\delta$ that is not constant, but decreases with $n$. We may, for example, take $\delta = \delta_1/\sqrt{n}$. One benefit of doing this is that the upper bound on $\delta$ in the theorem goes away and is replaced by a lower bound on $n$. Thus in the case of $N_1$ the theorem would require

$$n > \max\left( \left( \frac{c\,\mu_3\,\ell^d}{\chi(\rho_1^2\,\delta_1^2|\ell)} \right)^2, \left( \frac{4\,\Lambda^*\delta_1}{\vartheta^2} \right)^2, \left( \frac{\delta_1}{\delta_0} \right)^2 \right),$$

$$0.5\,\vartheta^2\,n - (2\,\Lambda^*\delta_1 + \vartheta\,s_1)\,\sqrt{n} + \ln\!\left( \chi(\rho_1^2\,\delta_1^2|\ell) - c\,\mu_3\,\ell^d/\sqrt{n} \right) \geqslant \ln(1/\varepsilon) - 0.5\,s_1^2.$$

The last part of the lower bound on $n$ is because we want $\forall n, \ \delta_1/\sqrt{n} < \delta_0$, for some $\delta_0 < 1$.

In the following subsections we present the results on which Theorems 4.1 and 4.4 are founded. One of these results is a lower bound on the probability of the set $F_n \cap \mathcal{C}(\delta)$ in Lemma 4.10; this is derived from (recent developments regarding) the multivariate Berry-Esseen Central Limit Theorem. Another is a lower bound on the probability of the set $\|f - \varphi^*\| \leqslant \vartheta$ in Lemma 4.14; this is based on (classical) concentration of measure inequalities. But before getting into these bounds, we point out certain properties of the MaxEnt p.d. that we will need.

## 4.2 Number of realizations, and probability under MaxEnt

We first review some properties of the p.d. that maximizes the entropy under linear equality and inequality constraints. Then we point out that in the presence of just equality constraints subject to tolerances, ratios of probabilities under the MaxEnt p.d. translate to ratios of numbers of realizations via multiplication by a factor which is close to 1 if the tolerance $\delta$ is small.

### 4.2.1 Equalities, inequalities, tolerances, and probabilities

With the background of §2.3, we can state the following concerning the effect of equalities, the two kinds of inequalities, and the presence or absence of tolerances, on the probabilities assigned by the MaxEnt p.d. to sequences whose frequency vectors satisfy the constraints.

**Proposition 4.6.**

1. *Suppose that the constraints ( 2.1) consist only of equalities and possibly inequalities that are not binding at $\varphi^*$, and there are no tolerances ($\delta = 0$). Then $\varphi^*$ assigns the same probability to all $n$-sequences whose frequency vectors satisfy the constraints; that common probability is $e^{-nH^*}$.*

2. *Suppose that the constraints are as above, but there is a tolerance $\delta > 0$. Then if $s, s'$ are two $n$-sequences with frequency vectors $f, f' \in \mathcal{C}(\delta)$, the ratio of their probabilities satisfies*

   $$e^{-2\Lambda^*\delta n} \leqslant \frac{\Pr_{\varphi^*}(s)}{\Pr_{\varphi^*}(s')} \leqslant e^{2\Lambda^*\delta n}$$

   *where $\Lambda^* = |\lambda^E| \cdot \beta^E$ has been defined in Theorem 4.1.*

3. *In the general case, with any kind of linear constraint and tolerance $\delta > 0$, we have*

   $$e^{-n\zeta} \leqslant \frac{\Pr_{\varphi^*}(s)}{\Pr_{\varphi^*}(s')} \leqslant e^{n\zeta}, \quad where \quad \zeta = 2\,(|\lambda^E| \cdot \beta^E)\,\delta + (\lambda^{\mathrm{BI}} \cdot \beta^{\mathrm{BI}})\,\delta + \Delta(\mathcal{C}(\delta)),$$

and where $\Delta(\mathcal{C}(\delta))$ is the solution to a linear program:

$$\Delta(\mathcal{C}(\delta)) \triangleq \max_{x \in \mathcal{C}'(\delta)} \sum_{1 \leqslant i \leqslant m} \lambda_i^{\mathrm{BI}} (b_i^{\mathrm{BI}} - A_{i.}^{\mathrm{BI}} \cdot x) \geqslant 0, \tag{4.5}$$

where $\mathcal{C}'(\delta)$ is $\mathcal{C}(\delta)$ with the additional constraints $x \succcurlyeq 0$, $\sum_i x_i = 1$.

4. *The exponent $\zeta$ above can also be found without involving the Lagrange multipliers, as the solution of the linear program*

$$\zeta = \max_{x,y \in \mathcal{C}'(\delta)} \sum_{1 \leqslant i \leqslant m} (x_i - y_i) \ln \varphi_i^*. \tag{4.6}$$

*This can also be written as $\max_{x \in \mathcal{C}'(\delta)} \sum_i x_i \ln \varphi_i^* - \min_{x \in \mathcal{C}'(\delta)} \sum_i x_i \ln \varphi_i^*$. It then follows that $\zeta \geqslant 0$.*

It can be seen that the result for case 2 follows from those for cases 1 and 3. The importance of the quantity $\Delta(\mathcal{C}(\delta))$ is that

1. If there are no inequalities binding at $\varphi^*$, $\Delta(\mathcal{C}(\delta)) \equiv 0$.

2. If there are binding inequalities, $\Delta(\mathcal{C}(\delta))$ is generally *non-zero* for $\delta = 0$. One way to look at $\Delta(\mathcal{C}(0))$ is this: it is known that if we have only linear equality constraints, any $f$ satisfying them is s.t. $H^* - H(f) = D(f \| \varphi^*)$; see e.g. [Csi96]. $\Delta(\mathcal{C}(0))$ expresses by how much $D(f \| \varphi^*)$ can deviate from $H^* - H(f)$ due *just* to the presence of binding inequalities in the constraints, irrespective of any tolerances.

3. The objective function in (4.5) depends only on the binding inequalities, whereas the domain over which it is maximized depends on *all* of the constraints, including non-binding inequalities, plus the tolerance $\delta$.

4. Recalling (2.2), it can be seen that (4.5) and (4.6) are parametric linear programs with parameter $\delta$. It is known that their solution is a *piecewise-linear, concave* function of $\delta$; however, finding this functional form exactly is non-trivial ([AM92]). A simple upper bound can be found by considering the dual LP, and this upper bound is linear in $\delta$.

The conclusion from Proposition 4.6 is that when the constraints are only equalities and inequalities non-binding at $\varphi^*$, the log ratio of the probabilities of sequences satisfying the constraints is bounded by a simple linear function of $n$ and the tolerance $\delta$, which, most importantly, is 0 when $\delta = 0$. In the presence of binding inequalities, whereas the linearity still obtains, the crucial difficulty is that even when the tolerance is 0, $n$ appears in the bound with a non-zero coefficient.

Now if $X, Y$ are two subsets of $F_n$, it follows from part 2 of Proposition 4.6 that the ratio of their probabilities under $\varphi^*$ and the ratio of their numbers of realizations are related by

$$e^{-2\Lambda^* \delta n} \leqslant \frac{\#X}{\#Y} \bigg/ \frac{\mathrm{Pr}_{\varphi^*}(X)}{\mathrm{Pr}_{\varphi^*}(Y)} \leqslant e^{2\Lambda^* \delta n}. \tag{4.7}$$

[This is a consequence of $\mathrm{Pr}_{\varphi^*}(X) = \sum_{\nu \in X} \#\nu \, \mathrm{Pr}_{\varphi^*}(\nu)$. In §4.1 we made use of (4.7) to derive Theorem 4.1 from (4.2).]

Another conclusion that can be drawn from Proposition 4.6 concerns the effect of *replacing* a constraint that turns out to be a binding inequality with an equality. Whereas this leaves the MaxEnt vector $\varphi^*$ unchanged, the set $\mathcal{C}'(\delta)$ becomes smaller and it follows from (4.6) that $\zeta$ will generally decrease (cannot increase). Hence the bounds on the ratio of probabilities in part 2 will generally be tighter, and so will the bounds in (4.7). In other words, *the solution will remain the same but the concentration around it will increase*, as intuitively expected.

Loosely speaking, one might say that the concentration phenomenon adds a 'variance' to the 'mean' provided by the MaxEnt solution; this is also demonstrated in the next example.

**Example 4.7.** Consider the case of $n$ die tosses with mean known to equal 4.5 and the additional knowledge that $f_1 + f_2 = 0.1$. Then we have $\varphi^* = (0.04339, 0.05661, 0.1445, 0.1885, 0.246, 0.321)$, and the vector of Lagrange multipliers corresponding to equalities is $\lambda^E = (-0.26598, 0.67117)$. Thus $2\Lambda^* \delta = 2.528\,\delta$ and $\Delta(\mathcal{C}(\delta)) \equiv 0$. The upper bound on $\delta$ in Theorem 4.4 is $\delta < \vartheta^2/5.056$. By part 2 of Proposition 4.6, any two sequences of $n$ tosses which satisfy the constraints have probabilities whose ratio lies in the interval $[e^{-2.53\delta n}, e^{2.53\delta n}]$.

Now suppose that instead of $f_1 + f_2 = 0.1$ we know that $f_1 + f_2 \leqslant 0.1$. This constraint turns out to be binding at the $\varphi^*$ we computed above, so the MaxEnt p.d. is the same as when we knew $f_1 + f_2 = 0.1$. However, the concentration around $\varphi^*$ is quite different. The Lagrange multipliers are $\lambda^E = -0.26598$, $\lambda^{\mathrm{BI}} = 0.67117$, and using part 3 of the proposition we find

$$\Delta(\mathcal{C}(0)) = 0.67117 \max_{x \in \mathcal{C}(0)} (0.1 - (1, 1, 0, 0, 0, 0) \cdot x) = 0.067117.$$

Consequently, any two sequences of $n$ tosses which satisfy these constraints have probabilities whose ratio lies in an interval never narrower than $[e^{-0.067n}, e^{0.067n}]$, irrespective of the tolerance $\delta$.

Finally, suppose that we know $\sum_i i\, f_i = 4.5$, $f_1 + f_2 \leqslant 0.1$, and $f_1 \geqslant 0.043$, $f_2 \geqslant 0.056$. The last constraint is non-binding at the $\varphi^*$ computed above, so $\varphi^*$ remains the same. However the new constraint affects the concentration, and $\Delta(\mathcal{C}(0))$ is now only $0.00067$.

### 4.2.2 Counting interpretation

When $m > n$, can we still give our results a counting interpretation, e.g. in terms of assigning $n$ balls to $m$ bins? The number of possible assignments, viewed as sequences of bin labels, is still $m^n$, but now any count vector $\nu$ will have at least $m - n$ of its elements $\nu_i$ equal to 0. If $\nu$ has $\mu$ non-zero elements, where we use $\mu$ by analogy to (3.6), the number of its realizations is still given by a multinomial coefficient: $\binom{n}{\nu_1, \dots, \nu_\mu}$, where $\mu \leqslant n$ and $\sum_i \nu_i = n$. See Example 4.15 in §4.4.

## 4.3 Lower bound on $\mathrm{Pr}_{\varphi^*}(F_n \cap \mathcal{C}(\delta))$

To bound the probability $\mathrm{Pr}_{\varphi^*}(F_n \cap \mathcal{C}(\delta))$, consider random variables $u_i$ taking values in the rows of the matrix $A = [A^E, A^{\mathrm{BI}}, A^{\mathrm{NI}}]^T$. The basic idea, taken from [Gr8], is that linear constraints on *counts* are expressible as constraints on the *sum* of a sequence of i.i.d. random variables which take values in the *set of coefficients* of the constraints. The probability of the set in which this sum lies can then be assessed via the Central Limit theorem. This probability will be significant if the distribution of the constructed random variables is such that their *means* are close to the *right-hand sides* of the constraints (the elements of the vector $b$).

To begin with, we define one $u_i$ for each row of $A^E$. $A^I$ needs a little more care because if our constraints include a two-sided inequality such as $c_1 \leqslant x_1 + \dots + \alpha_m\, x_m \leqslant c_2$, $A^I$ in (2.1) will contain the pair $(-\alpha_1, \dots, -\alpha_m)$, $(\alpha_1, \dots, \alpha_m)$ of dependent rows. Such dependencies have to be avoided for reasons that will be seen later (Proposition 4.9). Let $\ell_{I2} \geqslant 0$ be the number of such pairs of rows, or two-sided inequalities in $A^I$. For each of them either both 1-sided inequalities will be non-binding at $\varphi^*$, or one will be binding and the other non-binding. We therefore define a total of $\ell = \ell_E + \ell_I - \ell_{I2}$ random variables $u_1, \dots, u_\ell$ and collect them into a random vector $\boldsymbol{u} \in \mathbb{R}^\ell$. $u_i$ takes values in the $i$th row of the 'reduced' matrix

$$A' \triangleq [A^E, A^{\mathrm{BI2}}, A^{\mathrm{NI2}}]^T,$$

where $A^{\mathrm{BI2}}$, $A^{\mathrm{NI2}}$ are the reduced forms of $A^{\mathrm{BI}}$, $A^{\mathrm{NI}}$, with only one row $(\alpha_1, ..., \alpha_m)$ per two-sided inequality. (Example 4.8 below will make this clear.) Let all $u_i$ have the MaxEnt p.d. $\varphi^*$, i.e.

$$\Pr(u_i = a'_{ij}) = \varphi^*_j, \quad j = 1, ..., m, \tag{4.8}$$

irrespective of $i$. So all elements of $\boldsymbol{u}$ have the same p.d. but range over different sets, the rows of $A'$, and they are not independent. If $\boldsymbol{u}_1, ..., \boldsymbol{u}_n$ are i.i.d. with the p.d. of $\boldsymbol{u}$, the sum $\boldsymbol{u}_1 + \cdots + \boldsymbol{u}_n$ takes values of the form $\nu_1 A'_{\cdot 1} + \cdots + \nu_m A'_{\cdot m}$, with $\sum_j \nu_j = n$. Further, the p.d. $\varphi^*$ generates an $n$-sequence with frequency vector $f$ iff

$$\frac{\boldsymbol{u}_1 + \cdots + \boldsymbol{u}_n}{n} = f_1 A'_{\cdot 1} + \cdots + f_m A'_{\cdot m} = A' f,$$

hence by (2.2)

$$b^E - \delta \beta^E \preccurlyeq \frac{\boldsymbol{u}_1 + \cdots + \boldsymbol{u}_n}{n} \preccurlyeq b^E + \delta \beta^E, \quad \frac{\boldsymbol{u}_1 + \cdots + \boldsymbol{u}_n}{n} \preccurlyeq b^I + \delta \beta^I \quad \Longleftrightarrow \quad f \in F_n \cap \mathcal{C}(\delta). \tag{4.9}$$

By (4.8) all of $\boldsymbol{u}_1, ..., \boldsymbol{u}_n$ have mean $\bar{\boldsymbol{u}}$ and covariance matrix $\Sigma$ given by

$$\bar{u}_i = \sum_k a'_{ik} \varphi^*_k, \qquad \sigma_{ij} = \sum_k a'_{ik} a'_{jk} \varphi^*_k - \bar{u}_i \bar{u}_j. \tag{4.10}$$

From this expression for $\bar{u}_i$ we have $\bar{u}_i = b^E_i$ and $\bar{u}_i = b^{\mathrm{BI}}_i$ for the rows of $A^E$ and $A^{\mathrm{BI}}$, respectively, whereas for the rows of $A^{\mathrm{NI}}$, $\bar{u}_i = b^{\mathrm{NI}}_i - s^*_i$, with the slack $s^* > 0$. Therefore from (4.9)

$$f \in F_n \cap \mathcal{C}(\delta) \quad \Leftrightarrow \quad \frac{(\boldsymbol{u}_1 - \bar{\boldsymbol{u}}) + \cdots + (\boldsymbol{u}_n - \bar{\boldsymbol{u}})}{n} \in \mathbb{I}(\delta), \tag{4.11}$$

where $\mathbb{I}(\delta)$ is the $\ell$-dimensional rectangle $\mathbb{I}_1(\delta) \times \cdots \times \mathbb{I}_\ell(\delta)$ with elements

$$\mathbb{I}_i(\delta) = \begin{cases} [-\delta \beta^E_i, \delta \beta^E_i], & \text{equality,} \\ (-\infty, \delta \beta^I_i], & \text{binding inequality,} \\ (-\infty, s^*_i + \delta \beta^I_i], & \text{non-binding inequality,} \\ [-\delta \beta^I_i, s^*_{i+1} + \delta \beta^I_{i+1}], & \text{binding, non-binding pair,} \\ [-s^*_i - \delta \beta^I_i, \delta \beta^I_{i+1}], & \text{non-binding, binding pair,} \\ [-s^*_i - \delta \beta^I_i, s^*_{i+1} + \delta \beta^I_{i+1}], & \text{non-binding, non-binding pair,} \end{cases} \tag{4.12}$$

and where $\beta^E$, $\beta^I$ are the allowable error vectors appearing in the definition (2.2) of the set $\mathcal{C}(\delta)$.

**Example 4.8.** We want to find the probability density of a r.v. $v$ defined on the set $\{a_1, ..., a_{100}\} = \{-1, -0.98, ..., 0.98, 1\}$. The p.d. $x$ has to satisfy the constraints $\Pr(v > 0.95) = 0.02$, $E(v) \in [-0.1, 0.1]$, $E(v^2) \in [0.5, 0.6]$, $E(3 v^3 - 2 v) \in [-0.3, -0.2]$, and $\Pr(v < 0) \in [0.3, 0.4]$. (With the exception of the first constraint, the rest are taken from Example 7.2 of [BV04].)

For the equality constraints we have

$$A^E = [\; 0 \;\; ... \;\; 0 \;\; a_{98} \;\; a_{99} \;\; a_{100} \;], \quad b^E = [0.02],$$

and after $\varphi^*$ is found, with $H^* = 4.376$, we can classify the inequality constraints as

$$A^I = \begin{bmatrix} -a_1 & & ... & & -a_{100} \\ a_1 & & ... & & a_{100} \\ -a_1^2 & & ... & & -a_{100}^2 \\ a_1^2 & & ... & & a_{100}^2 \\ -(3 a_1^3 - 2 a_1) & & ... & (3 a_{100}^3 - 2 a_{100}) \\ 3 a_1^3 - 2 a_1 & & ... & 3 a_{100}^3 - 2 a_{100} \\ -a_1 & ... & -a_{50} & 0 & ... & 0 \\ a_1 & ... & a_{50} & 0 & ... & 0 \end{bmatrix}, \quad b^I = \begin{bmatrix} 0.1 \\ 0.1 \\ -0.5 \\ 0.6 \\ 0.3 \\ -0.2 \\ -0.3 \\ 0.4 \end{bmatrix}, \quad s^* = \begin{bmatrix} 0.1594 \\ 0.0406 \\ 6.8 \cdot 10^{-13} \\ 0.1 \\ 0.1 \\ 4.7 \cdot 10^{-13} \\ 0.1 \\ 4.4 \cdot 10^{-10} \end{bmatrix} \begin{matrix} \mathrm{NI} \\ \mathrm{NI} \\ \mathrm{BI} \\ \mathrm{NI} \\ \mathrm{NI} \\ \mathrm{BI} \\ \mathrm{NI} \\ \mathrm{BI} \end{matrix}.$$

Since $\ell_E = 1, \ell_I = 8$ and $\ell_{I2} = 4$, we define $\boldsymbol{u} = (u_1, ..., u_5)$. By (4.10) the mean of $\boldsymbol{u}$ is

$$\bar{\boldsymbol{u}} = A' \varphi^* = [A^E, A^I_{2.}, A^I_{4.}, A^I_{6.}, A^I_{8.}]^T \varphi^* = (0.02, 0.0594, 0.5, -0.2, 0.4),$$

and its covariance matrix is

$$\Sigma = \begin{bmatrix} 0.020 & 0.018 & 0.009 & 0.021 & -0.008 \\ 0.018 & 0.496 & -0.057 & 0.066 & -0.320 \\ 0.009 & -0.057 & 0.101 & -0.006 & 0.049 \\ 0.021 & 0.066 & -0.006 & 0.220 & 0.014 \\ -0.008 & -0.320 & 0.049 & 0.014 & 0.240 \end{bmatrix}.$$

If we take $\delta = 0.01$ (4.12) yields

$$\mathbb{I}(0.01) = ([-0.0002, 0.0002], [-0.1644, 0.0456], [-0.005, 0.106], [-0.103, 0.002], [-0.103, 0.004]).$$

Now the Central Limit Theorem in $\mathbb{R}^\ell$ ([Fel71], Theorem 2, §VIII.4) says that as $n \to \infty$,

$$\mathrm{Pr}_{\varphi^*}\left( \frac{(\boldsymbol{u}_1 - \bar{\boldsymbol{u}}) + \cdots + (\boldsymbol{u}_n - \bar{\boldsymbol{u}})}{\sqrt{n}} \in \sqrt{n}\,\mathbb{I}(\delta) \right) \to \mathcal{N}(\boldsymbol{z} \in \sqrt{n}\,\mathbb{I}(\delta) | \boldsymbol{0}, \Sigma)$$

where $\mathcal{N}(\cdot | \boldsymbol{0}, \Sigma)$ is the $\ell$-dimensional normal probability measure with mean $\boldsymbol{0}$ and covariance matrix $\Sigma$. To proceed, we note that

**Proposition 4.9.** *The covariance matrix* $\Sigma$ *of* $\boldsymbol{u}$ *has a non-singular square root* $\Sigma^{1/2}$.

Thus the inverse of $\Sigma^{1/2}$ exists, and denoting it by $\Sigma^{-1/2}$ we may define a 'standardized' version $\boldsymbol{v} = \Sigma^{-1/2}(\boldsymbol{u} - \bar{\boldsymbol{u}})$ of $\boldsymbol{u}$ with mean $\boldsymbol{0}$ and identity covariance matrix: cov $(\Sigma^{-1/2}\boldsymbol{u} - \Sigma^{-1/2}\bar{\boldsymbol{u}}) = \Sigma^{-1/2}\Sigma(\Sigma^{-1/2})^T = I$. Therefore from (4.11)

$$\begin{aligned} \mathrm{Pr}_{\varphi^*}(F_n \cap \mathcal{C}(\delta)) &= \mathrm{Pr}_{\varphi^*}\left( \frac{(\boldsymbol{u}_1 - \bar{\boldsymbol{u}}) + \cdots + (\boldsymbol{u}_n - \bar{\boldsymbol{u}})}{\sqrt{n}} \in \sqrt{n}\,\mathbb{I}(\delta) \right) \\ &= \mathrm{Pr}_{\varphi^*}\left( \frac{\boldsymbol{v}_1 + \cdots + \boldsymbol{v}_n}{\sqrt{n}} \in \mathbb{P}_n(\delta) \right), \end{aligned} \quad (4.13)$$

where $\mathbb{P}_n(\delta)$ is the polytope

$$\mathbb{P}_n(\delta): \sqrt{n}\,\mathbb{I}_{.1}(\delta) \preccurlyeq \Sigma^{1/2}\,w \preccurlyeq \sqrt{n}\,\mathbb{I}_{.2}(\delta), \quad w \in \mathbb{R}^\ell, \quad (4.14)$$

and where we think of $\mathbb{I}(\delta)$ as a $\ell \times 2$ matrix with the intervals $\mathbb{I}_i(\delta)$ as rows. The Berry-Esseen multi-variate Central Limit Theorem[4.3] says that if the $\ell$-dimensional random vectors $\boldsymbol{v}_i$ are i.i.d. with mean $\boldsymbol{0}$ and identity covariance matrix, the probability that $(\boldsymbol{v}_1 + \cdots + \boldsymbol{v}_n)/\sqrt{n}$ lies in a convex subset of $\mathbb{R}^\ell$, such as $\mathbb{P}_n(\delta)$, is close to the probability that a standard normal $\ell$-dimensional random vector $\boldsymbol{z}$ is in this set, and puts a bound on the deviation:

$$\left| \mathrm{Pr}\left( \frac{\boldsymbol{v}_1 + \cdots + \boldsymbol{v}_n}{\sqrt{n}} \in \mathbb{P}_n(\delta) \right) - \mathcal{N}\left( \boldsymbol{z} \in \mathbb{P}_n(\delta) \right) \right| \leqslant \frac{c\,\mu_3\,\ell^d}{\sqrt{n}}, \quad (4.15)$$

where $\mathcal{N}(\cdot)$ is the standard normal $\ell$-dimensional measure, $\mu_3 = E\|\boldsymbol{v}\|^3$, the 3d moment of the Euclidean norm of $\boldsymbol{v}$, and $c, d$ are two constants, for which [Ben03] and [CF11] give the values $(400, 1/4)$ and $(115, 1/2)$, respectively[4.4]. Independence is required of the $\boldsymbol{v}_i$, but the elements of $\boldsymbol{v}$ itself can exhibit arbitrary dependence.

---

4.3. See [Fel71], XVI.5 for the more familiar one-dimensional Berry-Esseen CLT.

4.4. [CF11] does not require that the vectors be i.i.d., merely independent. Then $\mu_3$ becomes $(1/n)\sum_{1 \leqslant i \leqslant n} E\|\boldsymbol{v}_i\|^3$.

To obtain a lower bound on $\mathcal{N}(\boldsymbol{z} \in \mathbb{P}_n(\delta))$, consider the largest hypersphere centered at the origin and contained in $\mathbb{P}_n(\delta)$. Let this sphere be $\mathbb{S}_n(\delta)$, with radius $\rho_n(\delta)$. Since the distance from a point $p$ to a hyperplane $a \cdot x = b$ is $|p \cdot a - b| / \|a\|_2$, the radius of $\mathbb{S}_1(\delta)$ is $\rho_1(\delta) = \min_i \big( |\mathbb{I}_{i1}| / \big\| \Sigma_{i.}^{1/2} \big\|_2$, $|\mathbb{I}_{i2}| / \big\| \Sigma_{i.}^{1/2} \big\|_2 \big)$ and the radius of $\mathbb{S}_n(\delta)$ is $\sqrt{n} \, \rho_1(\delta)$; by (4.12), $\rho_1(\delta)$ is positive if $\delta > 0$. Finally, using the fact that the elements of $\boldsymbol{z}$ are independent 1-dimensional standard normal r.v.'s. we have

$$\mathcal{N}(\boldsymbol{z} \in \mathbb{S}_n(\delta)) = \frac{1}{(2\pi)^{\ell/2}} \int_{z_1^2 + \cdots + z_\ell^2 \leqslant \rho_1^2(\delta) n} e^{-(z_1^2 + \cdots + z_\ell^2)/2} \, dz_1 \cdots dz_\ell.$$

This integral represents the probability that the independent standard normal r.v.'s $z_i$ are such that $z_1^2 + \cdots + z_\ell^2 \leqslant (\rho_1(\delta)\sqrt{n})^2$, hence (see e.g. [AS72], §26.4)

$$\mathcal{N}(\boldsymbol{z} \in \mathbb{S}_n(\delta)) = \chi(\rho_1^2(\delta) \, n \, | \, \ell), \quad \text{where} \quad \chi(w|\ell) \triangleq \frac{1}{2^{\ell/2} \Gamma(\ell/2)} \int_0^w t^{\ell/2 - 1} e^{-t/2} \, dt \qquad (4.16)$$

is the chi-squared distribution (c.d.f.) with $\ell$ degrees of freedom. Therefore

$$\mathcal{N}(\boldsymbol{z} \in \mathbb{P}_n(\delta)) > \chi(\rho_1^2(\delta) \, n \, | \, \ell), \quad \text{where} \quad \rho_1(\delta) \triangleq \min_{1 \leqslant i \leqslant \ell} \big( |\mathbb{I}_{i1}| / \big\| \Sigma_{i.}^{1/2} \big\|_2, \ |\mathbb{I}_{i2}| / \big\| \Sigma_{i.}^{1/2} \big\|_2 \big). \qquad (4.17)$$

Next we need the constant $\mu_3$ in (4.15). Noting that by (4.8) the vector $\boldsymbol{u}$ takes the value $A'_{.j}$, the $j$th column of $A'$, with probability $\varphi_j^*$,

$$\begin{aligned}
\mu_3 &= E\|\boldsymbol{v}\|^3 = E\|\Sigma^{-1/2}(\boldsymbol{u} - \bar{\boldsymbol{u}})\|_2^3 \\
&= E\big( \big( \Sigma_{1.}^{-1/2} \cdot (\boldsymbol{u} - \bar{\boldsymbol{u}}) \big)^2 + \cdots + \big( \Sigma_{\ell.}^{-1/2} \cdot (\boldsymbol{u} - \bar{\boldsymbol{u}}) \big)^2 \big)^{3/2} \\
&= \sum_{1 \leqslant j \leqslant m} \big( \big( \Sigma_{1.}^{-1/2} \cdot (A'_{.j} - \bar{\boldsymbol{u}}) \big)^2 + \cdots + \big( \Sigma_{\ell.}^{-1/2} \cdot (A'_{.j} - \bar{\boldsymbol{u}}) \big)^2 \big)^{3/2} \varphi_j^*.
\end{aligned} \qquad (4.18)$$

We have now established

**Lemma 4.10.** *Under the* MaxEnt *p.d. $\varphi^*$ of §2.3 the probability of the set of $n$-sequences whose frequency vectors lie in the $\ell$-dimensional set $\mathcal{C}(\delta)$ of constraints with tolerance $\delta > 0$ defined in (2.2) is*

$$\Pr_{\varphi^*}(F_n \cap \mathcal{C}(\delta)) > \chi(\rho_1^2(\delta) \, n \, | \, \ell) - c \, \mu_3 \, \ell^d / \sqrt{n},$$

*where $\chi(\cdot | \cdot)$ is the chi-squared distribution of (4.16), the radius $\rho_1(\delta)$, a linear function of $\delta$, is defined in (4.17), the constant $\mu_3$ is calculated by (4.18), and the pair of constants $c, d$ can be chosen either as $(115, 1/2)$ or $(400, 1/4)$.*

Several remarks are in order here. First, convenient lower bounds on $\chi(w|\ell)$ follow from its relationship to the incomplete gamma function, $\chi(w|\ell) = \gamma(\ell/2, w/2) / \Gamma(\ell/2)$. For example, for $\ell = 1, 2$ we have $\chi(w|\ell) \geqslant (1 - e^{-w/2}) / \Gamma(\ell/2)$; for $\ell \geqslant 3$ the bound becomes a little more complicated. (See [AS72], §26.4.19 and [OLBC10], §8.10.)

Second, Lemma 4.10 shows that as $n \to \infty$ the probability of the set $F_n \cap \mathcal{C}(\delta)$ tends to 1. Just this also follows from Sanov's theorem, [CT06] Theorem 11.4.1.

Third, the result of the lemma is to be compared with eq. (11) in §4 of [Gr8]. That result is based on the 'local' CLT, i.e. the density form, and shows that if the constraints are only equalities and there are no tolerances, then $\Pr_{\varphi^*}(F_n \cap \mathcal{C}(0))$ tends to a constant as $n \to \infty$[4.5].

---

[4.5]. Because of the absence of tolerances, that result is valid only for values of $n$ for which $F_n \cap \mathcal{C}(0)$ is not empty.

Fourth, the chi-squared distribution appears in the original (asymptotic) derivation of the entropy concentration phenomenon by Jaynes, in the Appendix of [Jay83]. Jaynes points out that $H(\varphi^*) - H(x) \approx C \, \|x - \varphi^*\|_2^2$, where $C$ is a constant, when the norm is small. Then the chi-squared distribution appears as the result of approximating a sum over a sphere of radius $\|x - \varphi^*\|_2$ by an integral.

Finally, besides the explicit dependence of the bound of Lemma 4.10 on $\ell$, there is also an implicit dependence, via $\rho_1(\delta)$ and the $\ell$-dimensional rectangle $\mathbb{I}(\delta)$ of (4.12). Otherwise, a striking feature is *the absence of any dependence on $m$*. This implies that the result holds even in the realm $n \leqslant m$. Are there any limitations to this? The following is an indication that there may be an implicit dependence on $m$, via $\mu_3$.

**Remark 4.11.** Consider the 1-dimensional version of (4.15), in the form given in [Fel71], XVI.5, Theorem 1: there the r.h.s. of (4.15) becomes $C \, (\mu_3'/\sigma^3)/\sqrt{n}$, where $\mu_3'$ is the absolute 3d moment and $C$ is an absolute constant. Now let $u$ be a scalar $m$-valued r.v. taking values in the set $\{0, 1, ..., 2\,k\}$, i.e. $m = 2\,k + 1$. $u$ takes the central value $k$ with probability $1 - 2\,k\,p$, and each of the other values with probability $p$. The mean of $u$ is $k$, its variance is $\sigma^2 = p\,(2\,k^3/3 + k^2 + k/3)$, and its 3d absolute central moment is $\mu_3' = p\,k^2\,(k+1)^2/2$. Then, if $p$ is, roughly, less than $1/(2\,k^2)$, $u$ is such that $\mu_3' \geqslant \sqrt{m}\,\sigma^3$. But this condition requires $n > C^2\,m$ when the Berry-Esseen CLT is applied to $u_1 + \cdots + u_n$.

We illustrate Lemma 4.10 with two examples.

**Example 4.12.** Returning to Example 4.8, we find

$$\Sigma^{1/2} = \begin{bmatrix} 0.1317 & 0.0177 & 0.0245 & 0.0350 & -0.0107 \\ 0.0177 & 0.6157 & -0.0418 & 0.0759 & -0.3310 \\ 0.0245 & -0.0418 & 0.3100 & -0.0087 & 0.0539 \\ 0.0350 & 0.0759 & -0.0087 & 0.4585 & 0.0494 \\ -0.0107 & -0.3310 & 0.0539 & 0.0494 & 0.3535 \end{bmatrix}.$$

By (4.17), with $\delta = 0.01$, the largest sphere centered at the origin and contained in the polytope $\mathbb{P}_1(\delta)$ defined by $\mathbb{I}_1(\delta) \preccurlyeq \Sigma^{1/2}w \preccurlyeq \mathbb{I}_2(\delta)$ has radius $\rho_1(0.01) = 0.001429$. By (4.18), $\mu_3 = 0.2106$, so by Lemma 4.10 with $(c, d) = (115, 1/2)$,

$$\Pr_{\varphi^*}(F_n \cap \mathcal{C}(0.01)) > \chi(2.04 \cdot 10^{-6}\,n\,|5) - 54.16/\sqrt{n}.$$

For $n = 8 \cdot 10^5, 9 \cdot 10^5, 10^6$ the value of the bound is $0.0423, 0.0718, 0.1025$.

**Example 4.13.** Returning to Example 4.7 with the constraints $\sum_i i\,f_i = 4.5$ and $f_1 + f_2 \leqslant 0.1$, we have

$$A' = [A^E, A^{\mathrm{BI}}]^T = \begin{bmatrix} 1 & 2 & 3 & 4 & 5 & 6 \\ 1 & 1 & 0 & 0 & 0 & 0 \end{bmatrix}, \quad b = [b^E, b^{\mathrm{BI}}]^T = \begin{bmatrix} 4.5 \\ 0.1 \end{bmatrix},$$

and $\varphi^* = (0.0434, 0.0566, 0.1445, 0.1885, 0.246, 0.321)$. Thus $\boldsymbol{u} = (u_1, u_2)$ with $\bar{\boldsymbol{u}} = (4.5, 0.1)$, and

$$\Sigma = \begin{bmatrix} 2.0413 & -0.2934 \\ -0.2934 & 0.09 \end{bmatrix}, \quad \Sigma^{1/2} = \begin{bmatrix} 1.4177 & -0.1767 \\ -0.1767 & 0.2424 \end{bmatrix}.$$

Now suppose we consider two different $\delta$: 0.005 for the equality and 0 for the binding inequality. Then the radius $\rho_1$ is 0 so Lemma 4.10 does not apply to the polytope

$$\mathbb{P}_n(\delta): \quad \sqrt{n} \begin{bmatrix} -0.0225 \\ -\infty \end{bmatrix} \preccurlyeq \Sigma^{1/2}z \preccurlyeq \sqrt{n} \begin{bmatrix} 0.0225 \\ 0 \end{bmatrix}.$$

Nevertheless we can easily find by numerical integration that $\mathcal{N}(z \in \mathbb{P}_n(\delta)) = \mathcal{N}(z \in \sqrt{n} \; \mathbb{I}(\delta)|\mathbf{0}, \Sigma) = 0.0625, 0.2845, 0.4423, 0.4992$ for $n = 10, 50, 100, 200$, respectively, 0.5 being the limit as $n \to \infty$. This illustrates that, as remarked in §2.2 after (2.2), tolerances on inequalities are not strictly necessary, but more a matter of analytical convenience, i.e. allowing $\mathbb{P}_1(\delta)$ to contain a hypersphere centered at the origin.

## 4.4 Upper bound on $\mathrm{Pr}_{\varphi^*}(\| f - \varphi^* \| > \vartheta)$

We now show that under the MaxEnt p.d., the probability of the set of vectors close to $\varphi^*$ either in $\ell_1$ or $\ell_2$ norm differs from 1 by a term exponentially small in $n$; in the $\ell_1$ case the term depends implicitly on $m$, whereas in the $\ell_2$ case it is independent of $m$.

**Lemma 4.14.** *Let*
$$s_1(\varphi^*) \triangleq \sum_{1 \leqslant i \leqslant m} \sqrt{\varphi_i^*(1 - \varphi_i^*)}, \qquad s_2(\varphi^*) \triangleq \sum_{1 \leqslant i \leqslant m} (\varphi_i^*)^2.$$
*Then for any $\vartheta > 0$,*
$$\begin{aligned}
\mathrm{Pr}_{\varphi^*}(\| f - \varphi^* \|_1 > \vartheta) &\leqslant e^{-(\vartheta \sqrt{n} - s_1)^2/2}, \\
\mathrm{Pr}_{\varphi^*}(\| f - \varphi^* \|_2 > \vartheta) &\leqslant e^{-(\vartheta \sqrt{n} - \sqrt{1 - s_2})^2/2}.
\end{aligned}$$

In the $\ell_1$ norm bound the worst case occurs when $\varphi^*$ is uniform: then $s_1(\varphi^*)$ is maximum, equal to $\sqrt{m-1}$, so the exponent is positive only if $n > (m-1)/\vartheta^2$. Also recall Remark 4.11 on the possible dependence of the bound on $\mathrm{Pr}_{\varphi^*}(F_n \cap \mathcal{C}(\delta))$ on $m$. In the $\ell_2$ norm bound however, $s_2(\varphi^*) < 1$ always, so the exponent is *independent* of $m$ and is positive if $n > (1 - s_2)/\vartheta^2$.

This completes the exposition of the material on which Theorem 4.4 is based, and now we can give an example illustrating the theorem.

**Example 4.15.** We return to Example 4.13. The presence of the binding inequality $f_1 + f_2 \leqslant 0.1$ does not allow us to apply Theorem 4.4 to study the concentration. However, suppose we exploit the remark made after (4.7) and change the binding inequality into an equality; then the theorem is applicable. The solution $\varphi^*$ is unaffected by this change, and so are most of the other quantities involved in the theorem: the covariance matrix $\Sigma$, the moment $\mu_3$, the multipliers $\lambda^E$, and the characteristics $s_1, s_2$ of $\varphi^*$. We have
$$\mu_3 = 4.2778, \quad \lambda^E = (-0.26598, 0.67117), \quad s_1(\varphi^*) = 2.075, \quad s_2(\varphi^*) = 0.2250.$$
The only effect is on the polytope $\mathbb{P}_n(\delta)$ which becomes
$$\sqrt{n} \begin{bmatrix} -4.5 \\ -0.1 \end{bmatrix} \delta \preccurlyeq \Sigma^{1/2} z \preccurlyeq \sqrt{n} \begin{bmatrix} 4.5 \\ 0.1 \end{bmatrix} \delta,$$
and consequently the radius $\rho_1$ is
$$\rho_1 = \min\left( \frac{4.5}{\|(1.4177, -0.1767)\|_2}, \frac{0.1}{\|(-0.1767, 0.2424)\|_2} \right) \delta = 0.3333 \, \delta.$$
The upper bound on $\delta$ in the theorem is $\delta < \vartheta^2/5.056$. With $\delta = 5 \cdot 10^{-5}, \vartheta = 0.025, \varepsilon = 10^{-30}$ we find $N_1 = N_2 = 2.968 \cdot 10^8$, much larger than the $N = 3.41 \cdot 10^5$ of Table 3.1. The limiting factor here is $N_0$, the solution to $\chi(\rho_1^2(\delta) \, n \, |\ell) - c \, \mu_3 \, \ell^d / \sqrt{n} = 0$, which turns out to be $2.968 \cdot 10^8$.

Now consider the case of the single constraint $\sum_i i\, f_i = \mu$, but with $m = 10^5$. Then the simple lower bound (3.8) on the result of Theorem 3.15 implies that $N \geqslant 10^{15}/4$. What does Theorem 4.4 say? The MAXENT p.d. is geometric with mean $\mu$, and it is numerically indistinguishable from the distribution on $\{0, 1, ...., \infty\}$, $\varphi_j^* = \frac{1}{\mu+1}\left(\frac{\mu}{\mu+1}\right)^j$. Table 4.1 shows the various quantities appearing in the theorem and their numerical values for $\mu = 4.5$ and $\mu = 45$.

| Quantities in Theorem 4.4 | $\mu = 4.5$ | $\mu = 45$ |
|---|---|---|
| $\varphi_j^* = \frac{1}{\mu+1}\left(\frac{\mu}{\mu+1}\right)^j$, $j \geqslant 0$ | | |
| $\lambda^E = \ln(1 + 1/\mu)$ | 0.201 | 0.022 |
| $\bar{u} = \mu$ | 4.5 | 45 |
| $\sigma = \mu\,(\mu+1)$ | 24.75 | 2070 |
| $\mu_3 = \sum_{j \geqslant 0}\left((j-\mu)^2/\sigma\right)^{3/2}\varphi_j^*$ | 2.4193 | 2.4146 |
| $\rho_1(\delta) = \delta\,\sqrt{\mu/(\mu+1)}$ | $0.905\,\delta$ | $0.989\,\delta$ |
| $s_1(\varphi^*)$ | 4.3126 | 13.441 |
| $s_2(\varphi^*)$ | 0.1 | 0.0110 |

**Table 4.1.** Geometric MAXENT distribution with mean $\mu$ on $\{0, 1, ..., \infty\}$.

Table 4.2 gives results for some values of $\delta, \vartheta, \varepsilon$. They are much better than the $\Omega(10^{15})$ result of Theorem 3.15, and in fact the values remain practically unchanged *no matter how large $m$ becomes*.

| $\delta$ | $\vartheta$ | $\varepsilon$ | $N_0$ | $N_1$ | $N_2$ |
|---|---|---|---|---|---|
| $10^{-4}$ | 0.02 | $10^{-80}$ | $3.87 \cdot 10^6$ | $3.65 \cdot 10^7$ | $1.33 \cdot 10^7$ |
| $10^{-5}$ | | | $3.85 \cdot 10^7$ | $3.85 \cdot 10^7$ | $3.85 \cdot 10^7$ |
| | 0.01 | | | | |
| $10^{-6}$ | | | $3.85 \cdot 10^8$ | $3.85 \cdot 10^8$ | $3.85 \cdot 10^8$ |
| | 0.005 | | | | |

| $\delta$ | $\vartheta$ | $\varepsilon$ | $N_0$ | $N_1$ | $N_2$ |
|---|---|---|---|---|---|
| $10^{-4}$ | 0.02 | $10^{-80}$ | $3.54 \cdot 10^6$ | $1.82 \cdot 10^{10}$ | $2.49 \cdot 10^8$ |
| $10^{-5}$ | | | $3.52 \cdot 10^7$ | $3.52 \cdot 10^7$ | $3.52 \cdot 10^7$ |
| | 0.01 | | | | |
| $10^{-6}$ | | | $3.52 \cdot 10^8$ | $3.52 \cdot 10^8$ | $3.52 \cdot 10^8$ |
| | 0.005 | | | | |

**Table 4.2.** Results from Theorem 4.4 with $\mu = 4.5$ (left) and $\mu = 45$ (right).

# 5  Conclusion

The phenomenon of entropy concentration appears when a large number of units is allocated to containers subject to constraints that are linear functions of the numbers of units in each container: most allocations will result in frequency (normalized count) vectors with entropy close to that of the vector of maximum entropy that satisfies the constraints. Asymptotic proofs of this phenomenon are known, beginning with the work of E. T. Jaynes, but here we presented a formulation entirely devoid of probabilities and provided explicit bounds on how large the number of units $n$ must be for concentration to any desired degree to occur. Our formulation also deals with the fact that equality constraints cannot be satisfied exactly by rational frequencies, but only to some prescribed tolerances. In addition, our version of concentration is in terms of deviation from the maximum entropy vector, more useful that the usual deviation form the maximum entropy value. Our results are not simple closed forms, because we try to minimize the required $n$, but the computational procedures can be implemented easily with any modern computer algebra system. Because of its conceptual simplicity and minimality of assumptions, entropy concentration is a powerful justification of the widely-used discrete MAXENT method, whenever it is applicable, and we believe that our bounds strengthen it considerably: by removing all asymptotic considerations and introducing tolerances, we turn arguments that invoke the concentration phenomenon from "in principle" or "in the limit" to "in practice".

**Acknowledgments.** We thank the authors of the CVXOPT convex optimization package for making their code publicly available. Thanks to S. J. Montgomery-Smith for help with the example in Remark 4.11. We are also grateful to an anonymous reviewer of an earlier version of this paper, for the suggestion to include a simple reference result on entropy concentration.

# Appendix A   Proofs for §2 and §3

## Proof of Proposition 2.1

Consider the equality constraints first. Writing them as $|A^E x - b^E| \preccurlyeq \delta \beta^E$, we see that they will be satisfied if $\max_i |A^E x - b^E|_i \leqslant \delta \min_i \beta_i^E$, or $\|A^E x - b^E\|_\infty \leqslant \delta \beta_{\min}^E$. Now for any $x \in \mathbb{R}^m$, $A^E \varphi^* = b^E \Leftrightarrow A^E x - b^E = A^E (x - \varphi^*)$. Thus $\|A^E x - b^E\|_\infty = \|A^E (x - \varphi^*)\|_\infty$. But $\|A^E (x - \varphi^*)\|_\infty \leqslant \||A^E\||_\infty \|x - \varphi^*\|_\infty$, where the (rectangular) matrix norm $\||\cdot\||_\infty$ is defined as the largest of the $\ell_1$ norms of the rows[A.1]. Therefore, to ensure $\|A^E x - b^E\|_\infty \leqslant \delta \beta_{\min}^E$ it suffices to require that $\|x - \varphi^*\|_\infty \leqslant \delta \beta_{\min}^E / \||A^E\||_\infty$.

Turning to the inequality constraints, write them as $A^I (\varphi^* + x - \varphi^*) \preccurlyeq b^I + \delta \beta^I$, or $A^I \varphi^* - b^I \preccurlyeq A^I (\varphi^* - x) + \delta \beta^I$. Since $A^I \varphi^* - b^I \preccurlyeq 0$, this inequality will be satisfied if $A^I (x - \varphi^*) \preccurlyeq \delta \beta^I$. This will certainly hold if $\max_i (A^I (x - \varphi^*))_i \leqslant \delta \min_i \beta_i^I$, which is equivalent to $\|A^I (x - \varphi^*)\|_\infty \leqslant \delta \beta_{\min}^I$. In turn, this will hold if we require $\||A^I\||_\infty \|x - \varphi^*\|_\infty \leqslant \beta_{\min}^I$.

For both types of constraints the final condition is stronger than necessary, but more so in the case of inequalities.

## Proof of Proposition 2.3

The proof is based on calculating the probability assigned by $\varphi^*$ to an $n$-sequence with frequency vector $f$ and equating the result to the general expression for the probability assigned to an $n$-sequence with that frequency vector by an arbitrary p.d. $p$.

We first find an expression for the maximum entropy $H^*$ in terms of the Lagrange multipliers. By (2.5),

$$
\begin{aligned}
H^* &= -\sum_{1 \leqslant j \leqslant m} \varphi_j^* \ln \varphi_j^* &&= \sum_{1 \leqslant j \leqslant m} \varphi_j^* (\lambda_0 + \lambda \cdot A_{\cdot j}) \\
&= \lambda_0 + \sum_j \varphi_j^* (\lambda \cdot A_{\cdot j}) &&= \lambda_0 + \sum_i \lambda_i \left( \sum_j \varphi_j^* a_{ij} \right) \\
&= \lambda_0 + \sum_i (\lambda_i^E b_i^E + \lambda_i^{\mathrm{BI}} \cdot b_i^{\mathrm{BI}}) &&= \lambda_0 + \lambda^E \cdot b^E + \lambda^{\mathrm{BI}} \cdot b^{\mathrm{BI}},
\end{aligned} \tag{A.1}
$$

where the second to last equality follows from the fact that $\varphi^*$ is s.t. $\sum_j a_{ij} \varphi_j^* = b_i$ for equalities and binding inequalities.

Now if an arbitrary sequence $s$ has frequency vector $f$, by (2.5) its probability under $\varphi^*$ is

$$
\mathrm{Pr}_{\varphi^*}(s) = \exp\left( -n \sum_{1 \leqslant j \leqslant m} (\lambda_0 + \lambda^E \cdot A_{\cdot j}^E + \lambda^{\mathrm{BI}} \cdot A_{\cdot j}^{\mathrm{BI}}) f_j \right) = e^{-n\xi(f)}. \tag{A.2}
$$

The exponent $\xi(f)$ in (A.2) can be written as

$$
\xi(f) = \lambda_0 + \sum_i \lambda_i^E (A_{i\cdot}^E \cdot f) + \sum_i \lambda_i^{\mathrm{BI}} (A_{i\cdot}^{\mathrm{BI}} \cdot f), \tag{A.3}
$$

---

[A.1]. For any rectangular matrix $A$ and compatible vector $x$, $\|A x\|_\infty \leqslant \||A\||_\infty \|x\|_\infty$ holds because the l.h.s. is $\max_i |A_{i\cdot} x|$. This is $\leqslant \max_i \sum_j |a_{ij} x_j| \leqslant \max_i \|x\|_\infty \|A_{i\cdot}\|_1 = \|x\|_\infty \||A\||_\infty$.

where the first sum ranges over the equalities and the second over the binding inequalities.

If $f$ is in $\mathcal{C}(\delta)$, then $A_{i\cdot}^E \cdot f - b_i^E \in [-\delta\,\beta_i^E, \delta\,\beta_i^E]$, and $A_{i\cdot}^{\mathrm{BI}} \cdot f_i \leqslant b^{\mathrm{BI}} + \delta\,\beta_i^{\mathrm{BI}}$. But whereas $\lambda^{\mathrm{BI}} \geqslant 0$, the elements of $\lambda^E$ can be positive or negative. Therefore from (A.3)[A.2]

$$
\begin{aligned}
\max_{f \in \mathcal{C}(\delta)} \xi(f) &\leqslant \lambda_0 + \lambda^E \cdot b^E + (|\lambda^E| \cdot \beta^E)\,\delta + \lambda^{\mathrm{BI}} \cdot (b^{\mathrm{BI}} + \delta\,\beta^{\mathrm{BI}}), \\
\min_{f \in \mathcal{C}(\delta)} \xi(f) &\geqslant \lambda_0 + \lambda^E \cdot b^E - (|\lambda^E| \cdot \beta^E)\,\delta + \min_{f \in \mathcal{C}(\delta)} \sum_i \lambda_i^{\mathrm{BI}}(A_{i\cdot}^{\mathrm{BI}} \cdot f),
\end{aligned}
\tag{A.4}
$$

where $|\cdot|$ indicates element-wise absolute value. Substituting the last expression of (A.1) into (A.4) we get

$$
\begin{aligned}
\max_{f \in \mathcal{C}(\delta)} \xi(f) &\leqslant H^* + (|\lambda^E| \cdot \beta^E + \lambda^{\mathrm{BI}} \cdot \beta^{\mathrm{BI}})\,\delta, \\
\min_{f \in \mathcal{C}(\delta)} \xi(f) &\geqslant H^* - (|\lambda^E| \cdot \beta^E)\,\delta + \min_{f \in \mathcal{C}(\delta)} \sum_i \lambda_i^{\mathrm{BI}}(A_{i\cdot}^{\mathrm{BI}} \cdot f - b_i^{\mathrm{BI}}) \\
&= H^* - (|\lambda^E| \cdot \beta^E)\,\delta - \max_{f \in \mathcal{C}(\delta)} \sum_i \lambda_i^{\mathrm{BI}}(b_i^{\mathrm{BI}} - A_{i\cdot}^{\mathrm{BI}} \cdot f) \\
&= H^* - (|\lambda^E| \cdot \beta^E)\,\delta - \Delta(\mathcal{C}(\delta)).
\end{aligned}
\tag{A.5}
$$

Finally, for any p.d. $p$ and any $n$-sequence $s$ with frequency vector $f$, the probability of $s$ can be written as $\Pr_p(s|f) = e^{-n(H(f) + D(f\|p))}$. Comparing this with (A.2), and using (A.5) we find that for any $f \in \mathcal{C}(\delta)$

$$
-(|\lambda^E| \cdot \beta^E)\,\delta - \Delta(\mathcal{C}(\delta)) \leqslant H(f) + D(f\|\varphi^*) - H^* \leqslant (|\lambda^E| \cdot \beta^E + \lambda^{\mathrm{BI}} \cdot \beta^{\mathrm{BI}})\,\delta.
\tag{A.6}
$$

The claim of the proposition follows from the 2nd inequality above. We will use the full (A.6) in Proposition 4.6 of §4.2.

## Proof of Proposition 3.3

From the explanation after Definition 3.1, the adjustment of $\tilde{\nu}$ to $\nu^*$ ensures $\|\nu^* - n\,\varphi^*\|_\infty \leqslant 1$; this establishes the $\ell_\infty$ claim. More precisely, this adjustment causes $d$ of the elements of $\nu^*$ to differ from the corresponding elements of $n\,\varphi^*$ by $<1$, and the rest to differ by $\leqslant 1/2$, so $\|\nu^* - n\,\varphi^*\|_1 \leqslant \max_d (d + (m-d)/2)$, and since $d \leqslant \lfloor m/2 \rfloor$ this cannot exceed $3\,m/4$; this establishes the claim for the $\ell_1$ norm.

Now for the bound on $H(f^*)$ we use the result of Problem 3.10 in [CK11]: if $\|p - q\|_1 = \zeta$, then $|H(p) - H(q)| \leqslant (1/2)\,\zeta \ln(m-1) + h(\zeta/2)$, where $h(x) = -x \ln x - (1-x) \ln(1-x)$ is the binary entropy function[A.3]. The function $g(\zeta) = (1/2)\,\zeta \ln(m-1) + h(\zeta/2)$ increases with $\zeta$, so if $\|f^* - \varphi^*\|_1 \leqslant 3\,m/(4\,n)$, then $H^* - H(f^*) \leqslant g(3\,m/(4\,n))$. Thus $H(f^*) \geqslant H^* - \frac{3m}{8n} \ln \frac{8(m-1)n}{3m} + \left(1 - \frac{3m}{8n}\right) \ln\left(1 - \frac{3m}{8n}\right)$, from which the result in the proposition follows.

## Proof of Proposition 3.6

The first statement of the proposition follows from Theorem 3 of [HY10] which, for a given *arbitrary* p.d. $p$, shows how to construct a p.d. $q$ which minimizes $H(q)$ subject to $\|p - q\|_1 \leqslant \zeta$. Taking $p$ to be $\varphi^*$, it suffices to note that if $\zeta \leqslant \varphi_{\min}^*$, the $K$ in that theorem becomes our $m$. Then $q$ is found by putting a copy of $\varphi^*$ in decreasing order, adding $\zeta/2$ to its first element, and subtracting $\zeta/2$ from the last element. With this, $H(\varphi^*) - H(q)$ becomes $h_2(\varphi_{\max}^*, \varphi_{\min}^*, \zeta)$.

---

[A.2.] Here we use the following: suppose we know $-d \leqslant x \leqslant d$, where $d > 0$; then given some $a \in \mathbb{R}$, the most we can say about $a\,x$ is $-|a|\,d \leqslant a\,x \leqslant |a|\,d$, or $-d \leqslant (a/|a|)\,x \leqslant d$.

[A.3.] [CK11] use binary logs throughout, we use natural logs throughout.

To prove the second statement, Proposition 2.3 says that any $f \in F_n \cap \mathcal{C}(\delta)$ is s.t. $H(f) \leqslant H^* + \Lambda^* \delta - D(f \| \varphi^*)$. Then [OW05] Theorem 2.1 (see also Remark 3.4.6 in [RS14]), the distribution-dependent improvement to Pinsker's inequality, says that $D(f \| \varphi^*) \geqslant c(\varphi^*) \| f - \varphi^* \|_1^2$, and the desired result follows.

**Proof of Proposition 3.10**

For any $m$-vector $x$, $\|x\|_1 \leqslant \sqrt{m} \, \|x\|_2$. Therefore the set in the proposition is a superset of $\{f \in F_n, \|f - \varphi^*\|_2 \leqslant \vartheta / \sqrt{m}\}$, so it will suffice to put a lower bound on the size of this set. Since

$$f \in F_n, \|f - \varphi^*\|_2 \leqslant \vartheta / \sqrt{m} \quad \Longleftrightarrow \quad \nu \succcurlyeq 0, \sum_i \nu_i = n, \|\nu - n \varphi^*\|_2 \leqslant n \vartheta / \sqrt{m},$$

we consider the number of lattice points in the set

$$x \in \mathbb{R}^m, \ x \succcurlyeq 0, \ \sum_i x_i = n, \ \|x - n \varphi^*\|_2 \leqslant n \vartheta / \sqrt{m}. \tag{A.7}$$

This set is the intersection of an $m$-dimensional sphere with a hyperplane passing through its center, contained in the first orthant. The condition $\vartheta \leqslant \varphi_{\min}^* / \sqrt{m}$ ensures that the distance from the center $n \varphi^*$ of the sphere to the closest coordinate hyperplane is not less than the radius $n \vartheta / \sqrt{m}$ of the sphere. Hence the set (A.7) is an $(m-1)$-dimensional sphere of radius $n \vartheta / \sqrt{m}$ centered at $n \varphi^*$, wholly contained in the first orthant. By Proposition 3.8, if $m \geqslant 3$ and the radius $r = n \vartheta / \sqrt{m}$ is at least $\sqrt{m-1} / 2$, i.e. $n \vartheta \geqslant m / 2$, the number of lattice points in this sphere, hence in the set (A.7), and therefore in $\{f \in F_n, \|f - \varphi^*\|_1 \leqslant \vartheta\}$ as well, is at least

$$\frac{\pi^{(m-1)/2}}{\Gamma((m+1)/2)} \frac{(n \, \vartheta)^{m-1}}{m^{(m-1)/2}} \left(1 - \frac{\sqrt{m(m-1)}}{2 \, n \, \vartheta}\right)^{m-1},$$

and the last expression is at least $\left(1 - m/(2 \, n \, \vartheta)\right)^{m-1}$.

**Proof of Lemma 3.11**

$\#\mathcal{B}_n$ is the sum over all $f \in \mathcal{B}_n$ of $\# f$. Similarly to what [CT06] do in the proof of the conditional limit theorem, Theorem 11.6.2, we simply bound the sum over $F_n \cap \mathcal{C}(\delta)$ from above by the sum over all of $F_n$. [This is not as bad as it seems. We don't know of a simple bound substantially better than $|F_n| = O(n^{m-1})$. E.g. bounding $\mathcal{C}(\delta) \cap F_n$ by a sphere or hypercube would lead to an $O(n^{m-1})$ bound; compare, for example, with Proposition 3.10.] Then using (3.6) and Proposition 3.6 on $\# f$,

$$\# B_n(\delta, \vartheta) \leqslant \sum_{f \in F_n, \, \|f - \varphi^*\|_1 > \vartheta} \# f \leqslant e^{n(H^* + \Lambda^* \delta - c(\varphi^*) \vartheta^2)} \sum_{f \in F_n} S(f, \mu). \tag{A.8}$$

Now if $F_n^{(\mu)}$ is the subset of $F_n$ consisting of vectors with $\mu$ non-zero elements,

$$\sum_{f \in F_n} S(f, \mu) = \sum_{\mu=1}^{m} \sum_{f \in F_n^{(\mu)}} S(f, \mu) = \sum_{\mu=1}^{m} \binom{m}{\mu} \sum_{\substack{f_1 = \nu_1/n, \ldots, f_\mu = \nu_\mu/n \\ \nu_1 + \cdots + \nu_\mu = n, \, \nu_i \geqslant 1}} S(f, \mu),$$

where the $\binom{m}{\mu}$ comes from the fact that $\# f$ depends only on the non-zero elements and not on their positions. Thus by (3.6)

$$\sum_{f \in F_n} S(f, \mu) = \sum_{\mu=1}^{m} \binom{m}{\mu} \frac{1}{(2 \pi n)^{\frac{\mu-1}{2}}} (\sqrt{n})^\mu \sum_{\substack{\nu_1 + \cdots + \nu_\mu = n \\ \nu_1, \ldots, \nu_\mu \geqslant 1}} \frac{1}{\sqrt{\nu_1 \cdots \nu_\mu}}. \tag{A.9}$$

We now need an auxiliary result, proved in Appendix C, on the inner sum in (A.9): for any $\mu \geqslant 2$,

$$\sum_{\substack{\nu_1 + \cdots + \nu_\mu = n \\ \nu_1, \ldots, \nu_\mu \geqslant 1}} \frac{1}{\sqrt{\nu_1 \cdots \nu_\mu}} < \frac{\pi^{\mu/2}}{\Gamma(\mu/2)} \, n^{\mu/2 - 1}. \tag{A.10}$$

Using this in (A.9),

$$\sum_{f \in F_n} S(f, \mu) = \sqrt{2\pi/n} \sum_{\mu=1}^m \binom{m}{\mu} \left(\frac{n}{2}\right)^{\mu/2} \frac{1}{\Gamma(\mu/2)} < 4 \sqrt{2\pi/n} \left(1 + \sqrt{n/4}\right)^m,$$

where for the inequality we used $\Gamma(\mu/2) \geqslant 2^{\mu/2 - 2}$. Combining the above with (A.8), after a small manipulation we obtain the result of the lemma.

**Proof of Lemma 3.12**

Given an $\alpha \in (0, 1)$, $\mathcal{A}_n(\delta, \alpha\,\vartheta)$ is a subset of $\mathcal{A}_n(\delta, \vartheta)$. We put a lower bound on $\#\mathcal{A}_n(\delta, \alpha\,\vartheta)$ by first obtaining a lower bound on the size of $\mathcal{A}_n(\delta, \alpha\,\vartheta)$, and then putting a lower bound on the $\# f$ of the $f \in \mathcal{A}_n(\delta, \alpha\,\vartheta)$. The result is a lower bound on $\#\mathcal{A}_n(\delta, \vartheta)$.

By Proposition 3.10, if $n \geqslant m / (2\,\alpha\,\vartheta)$ and $\alpha\,\vartheta \leqslant \sqrt{m}\;\varphi^*_{\min}$, the set $\{f \in F_n, \|f - \varphi^*\|_1 \leqslant \alpha\,\vartheta\}$ contains at least $L(n, m, \alpha\,\vartheta)$ elements. By Proposition 2.1, if $\|f - \varphi^*\|_\infty \leqslant \delta\,\vartheta_\infty$ then $f \in \mathcal{C}(\delta)$. Therefore

$$n \geqslant \frac{m}{2\,\alpha\,\vartheta}, \quad \alpha\,\vartheta < \min\left(\sqrt{m}\;\varphi^*_{\min}, \delta\,\vartheta_\infty\right) \quad \Rightarrow \quad |\mathcal{A}_n(\delta, \alpha\,\vartheta)| \geqslant L(n, m, \alpha\,\vartheta). \tag{A.11}$$

Now consider the function $\sigma(f) = 1 / \sqrt{f_1 \cdots f_\mu}$, $f \succ 0$, and $\|f - \varphi^*\|_1 \leqslant \alpha\,\vartheta$. We put a lower bound on it ignoring the constraint $\sum_i f_i = 1$; the bound will hold a fortiori when the constraint is present. $\sigma(f)$ is convex, hence it lies above the hyperplane tangent to it at the point $\varphi^*$: $\sigma(f) \geqslant \sigma(\varphi^*) + \nabla\sigma(\varphi^*) \cdot (f - \varphi^*)$. [In this equation we are taking $f$ to be $(f_1, \ldots, f_\mu, 0, \ldots, 0)$]. Now $\nabla\sigma(f) = (-1/2)\,\sigma(f)\,(1/f_1, \ldots, 1/f_\mu)$, and $|\nabla\sigma(\varphi^*) \cdot (f - \varphi^*)| \leqslant \|\nabla\sigma(\varphi^*)\|_1 \|f - \varphi^*\|_1$. Since $\|f - \varphi^*\|_1 \leqslant \alpha\,\vartheta$, we have the bound

$$\sigma(f) \geqslant \sigma(\varphi^*) \left(1 - \frac{\alpha\,\vartheta}{2} \sum_i 1/\varphi^*_i\right).$$

Using (3.6), the above bound on $\sigma(f)$, and putting $\zeta = \alpha\,\vartheta$ in the first bound of Proposition 3.6, we see that for any $f \in \mathcal{A}_n(\delta, \alpha\,\vartheta)$,

$$\alpha\,\vartheta \leqslant \varphi^*_{\min} \quad \Rightarrow$$
$$\# f \geqslant \frac{e^{-\frac{\mu}{12}}}{(2\pi n)^{\frac{\mu-1}{2}} \sqrt{\varphi^*_1 \cdots \varphi^*_m}} \left(1 - \frac{\alpha\,\vartheta}{2} \sum_i 1/\varphi^*_i\right) e^{n(H^* - h_2(\varphi^*_{\max}, \varphi^*_{\min}, \alpha\,\vartheta))}. \tag{A.12}$$

Combining (A.11) with (A.12) and noting that the minimum of (A.12) w.r.t. $\mu$ occurs at $\mu = m$, we obtain the result of the lemma.

**Proof of Proposition 3.13**

Since the function $h_2$ decreases when $\alpha$ decreases (recall comment after Proposition 3.6), the *maximum* of $\psi$ w.r.t. $\alpha$ occurs at $\alpha = 0$; therefore we definitely need $\delta < \left(c(\varphi^*) / \Lambda^*\right) \vartheta^2$ if $\psi$ is to be positive.

Since $h_2$ increases with $\alpha$, $\alpha$ cannot be too large. To limit $\alpha$ we use an upper bound on $h_2$: writing $\zeta$ as $(1 - \zeta/\varphi^*_{\min})\,0 + (\zeta/\varphi^*_{\min})\,\varphi^*_{\min}$ it follows from the convexity of $h_2$ that

$$h_2(\varphi^*_{\max}, \varphi^*_{\min}, \zeta) \leqslant (\zeta/\varphi^*_{\min})\,h_2(\varphi^*_{\max}, \varphi^*_{\min}, \varphi^*_{\min}).$$

[The graph of $h_2(\zeta)$ lies below the line from the origin to the point $(\varphi_{\min}^*, h_2(\varphi_{\min}^*))$.] The upper bound on $\alpha$ follows by using this bound on $h_2$ in the condition $\psi(\delta, \vartheta, \alpha) > 0$.

**Proof of Theorem 3.14**

Using Lemmas 3.11 and 3.12 we find that

$$\frac{\#\mathcal{A}_n(\delta, \vartheta)}{\#\mathcal{B}_n(\delta, \vartheta)} \geqslant C_0(m) \left(1 - \frac{\alpha\,\vartheta}{2} \sum_{1 \leqslant i \leqslant m} 1/\varphi_i^*\right) \frac{(\alpha\,\vartheta - m/(2\,n))^{m-1}}{(1 + 2/\sqrt{n})^m} e^{n\psi(\delta, \vartheta, a)}, \tag{A.13}$$

where the function $\psi$ is as in Proposition 3.13, and, after some simplification, the constant $C_0$ is

$$C_0(m) = \frac{e^{-\frac{m}{12}} (m/2)^{-\frac{m-1}{2}}}{2\sqrt{2\,\pi}\,\Gamma((m+1)/2)\,\sqrt{\varphi_1^* \cdots \varphi_m^*}}.$$

Using the condition $n \geqslant m/(2\,\alpha\,\vartheta)$ of the lemma we can eliminate $n$ from the denominator of the third factor in (A.13) to get

$$\frac{\#\mathcal{A}_n(\delta, \vartheta)}{\#\mathcal{B}_n(\delta, \vartheta)} \geqslant C_0(m) \frac{1 - (\alpha\,\vartheta/2) \sum_{1 \leqslant i \leqslant m} 1/\varphi_i^*}{\left(1 + \sqrt{8\,\alpha\,\vartheta/m}\right)^m} \left(\alpha\,\vartheta - \frac{m}{2\,n}\right)^{m-1} e^{n\psi(\delta, \vartheta, a)},$$

and defining $C_1(m, \vartheta, \alpha)$ to be the 2nd factor on the right we get (3.7) in the theorem.

**Proof of Theorem 3.15**

Continuing from the proof of Theorem 3.14, to have $\#\mathcal{A}_n(\delta, \vartheta)/\#\mathcal{B}_n(\delta, \vartheta) \geqslant 1/\varepsilon$ we need

$$(m-1)\ln\left(\alpha\,\vartheta - m/(2\,n)\right) + \ln C_0(m)\,C_1(m, \vartheta, \alpha) + n\,\psi(\delta, \vartheta, \alpha) \geqslant \ln 1/\varepsilon.$$

Using the fact that $\ln(1 - x) > -x/(1 - x)$ for $x \in (0, 1)$, this will hold for all $n$ greater than the solution of

$$n\,\psi(\delta, \vartheta, \alpha) - \frac{m\,(m-1)/(2\,\alpha\,\vartheta)}{n - m/(2\,\alpha\,\vartheta)} = \lambda(m, \varepsilon, \vartheta, \alpha) \tag{A.14}$$

where

$$\lambda(m, \varepsilon, \vartheta, \alpha) \triangleq \ln 1/\varepsilon - \ln C_0(m)\,C_1(m, \vartheta, \alpha) + (m-1)\ln 1/(\alpha\,\vartheta). \tag{A.15}$$

So $n$ must satisfy the conditions $n \geqslant m/(2\,\alpha\,\vartheta)$ and (A.14), both of which depend on the yet-to-be-specified parameter $\alpha$. (A.14) is a quadratic in $n$, and its roots are (using abbreviated notation)

$$\frac{m\,\psi + 2\,\alpha\,\vartheta\,\lambda \pm \sqrt{(m\,\psi - 2\,\alpha\,\vartheta\,\lambda)^2 + 8\,\alpha\,\vartheta\,m\,(m-1)\,\psi}}{4\,\alpha\,\vartheta\,\psi}.$$

It can be seen that no matter what is the sign of $m\,\psi - 2\,\alpha\,\vartheta\,\lambda$, the condition $n \geqslant m/2\,\alpha\,\vartheta$ excludes the solution with the minus in front of the square root. Using the inequality $\sqrt{x^2 + y^2} < |x| + |y|$ we can put an upper bound on the other solution, and after considering the two cases for $|m\,\psi - 2\,\alpha\,\vartheta\,\lambda|$ we arrive at

$$N(\varepsilon, \vartheta, \alpha) = \max\left(\frac{m}{2\,\alpha\,\vartheta}, \frac{\lambda}{\psi}\right) + \frac{m}{\sqrt{2\,\alpha\,\vartheta\,\psi}}. \tag{A.16}$$

Finally, the defitnition of $\hat{\alpha}$ collects the upper bounds on $\alpha$ in Lemma 3.12 and Proposition 3.13.

**Proof of Theorem 3.17**

By Proposition 2.1 the first condition on $n$ ensures that $f^*$ is in $\mathcal{C}(\delta)$, and the second that the set $\|f - \varphi^*\|_1 \leqslant 3\,m/4\,n$ to which $f^*$ belongs and the set $\|f - \varphi^*\|_1 > \vartheta$ are disjoint.

Using (3.6) to put a lower bound on $\#f^*$ and Lemma 3.11 to put an upper bound on $\#\mathcal{B}(\delta,\vartheta)$, and noting that $H(f^*) - H^*$ is lower-bounded by the last result of Proposition 3.3, we have

$$\frac{\#f^*}{\#\mathcal{B}_n(\delta,\vartheta)} \geqslant \frac{e^{-m/12}(1-3\,m/8\,n)^{n-3m/8}}{4\,(2\,\pi)^{m/2}\,n^{m-1}\,(8\,n/3)^{3m/8}\,\sqrt{f_1^*\cdots f_m^*}\,(1/2+1/\sqrt{n})^m}\,e^{n(c(\varphi^*)\vartheta^2-\Lambda^*\delta)}. \tag{A.17}$$

Now we make the simplifications

$$(1-3\,m/8\,n)^{n-3m/8} \;>\; e^{-3m/8},$$
$$(1/2+1/\sqrt{n})^m \;<\; (1/2)^m\,(1+\sqrt{2}\,\vartheta)^m,$$
$$4\,(2\,\pi)^{m/2}\,(8/3)^{3m/8} \;<\; 4^{m+1}.$$

The first follows from the fact that $\ln(1-x) > -x\,/\,(1-x)$ for $x \in (0,\,1)$, and the second from assuming that $n > 2/\vartheta^2$. Using these inequalities in the r.h.s. of (A.17),

$$\frac{\#f^*}{\#\mathcal{B}_n(\delta,\vartheta)} > \frac{2^{-m-2}\,e^{-m/2}}{\sqrt{f_1^*\cdots f_m^*}\,(1+\sqrt{2}\,\vartheta)^m}\,\frac{e^{n(c(\varphi^*)\vartheta^2-\Lambda^*\delta)}}{n^{1.375m-1}} = C(f^*,\vartheta,m)\,\frac{e^{n(c(\varphi^*)\vartheta^2-\Lambda^*\delta)}}{n^{1.375m-1}},$$

and we want this to be $\geqslant 1/\varepsilon$, i.e.

$$n\,\big(c(\varphi^*)\,\vartheta^2 - \Lambda^*\,\delta\big) - (1.375\,m-1)\ln n \geqslant \ln(1/\varepsilon) + \ln(1/C). \tag{A.18}$$

But $\ln(1/C) < m\,\big(\ln 2 + 1/2 + \sqrt{2}\,\vartheta\big) + 1/2\sum_i \ln f_i^* + 2\ln 2$, and the last condition on $n$ in the theorem follows by using this in (A.18). This condition subsumes the assumption that $n > 2/\vartheta^2$.

**Proof of Corollary 3.18**

It can be seen that when $\varphi^* = (1/m, ..., 1/m)$, the count vector $\nu^*$ of Definition 3.1 is such that $n \bmod m$ of its elements are equal to $\lceil n/m \rceil$ and the rest are equal to $\lfloor n/m \rfloor$. Therefore the frequency vector $f^* = u^*$ is such that

$$\ln f_i^* \leqslant \ln(\lceil n/m \rceil/n) \leqslant \ln((n/m+1)/n) = \ln(1/m+1/n) \leqslant \ln(2/m).$$

Thus the 3d condition on $n$ in Theorem 3.17, where here we have $\delta \equiv 0$, $c(\varphi^*) = 1/2$, will hold if

$$n \geqslant \frac{(2.75\,m-2)\ln n + (2.4+3\,\vartheta)\,m + m\ln(2/m) + 2\ln(4/\varepsilon)}{\vartheta^2},$$

and this is implied by the 2nd condition on $n$ given in the Corollary.

# Appendix B  Proofs for §4

**Proof of Proposition 4.6**

From (A.5) and (A.6) in the proof of Proposition 2.3 we observe the following:

1. Suppose that the constraints are only equalities and, possibly, non-binding inequalities, and the tolerance $\delta$ is 0. Then $\Delta(\mathcal{C}(0)) \equiv 0$ and by (A.5), $\max \xi = \min \xi = H^*$. So by (A.2) the probability of any sequence whose frequency vector satisfies the constraints is

$$\Pr_{\varphi^*}(s) = e^{-nH^*}.$$

Since $\Delta(\mathcal{C}(0)) \equiv 0$, (A.6) shows that $H^* - H(f) = D(f\,\|\,\varphi^*)$.

2. Still with $\delta = 0$, (A.6) shows that in the presence of inequality constraints, $\varphi^*$ assigns to frequency vectors that satisfy the constraints probabilities that may *exceed* $e^{-nH^*}$. (Still, $f$ with entropies less than $H^*$ have far fewer realizations.)

3. Given two sequences $s$ and $s'$ whose frequency vectors $f$ and $f'$ are in $\mathcal{C}(\delta)$, it follows from (A.2) and (A.5) or (A.6) that

$$e^{-n\zeta} \leqslant \frac{\Pr_{\varphi^*}(s)}{\Pr_{\varphi^*}(s')} \leqslant e^{n\zeta}, \quad \text{where} \quad \zeta = \left(2\,|\lambda^E| \cdot \beta^E + \lambda^{\mathrm{BI}} \cdot \beta^{\mathrm{BI}}\right)\delta + \Delta(\mathcal{C}(\delta)).$$

4. If the sequences $s$, $s'$ have frequency vectors $f$, $f'$ respectively, then $\Pr_{\varphi^*}(s) / \Pr_{\varphi^*}(s') = \prod_i (\varphi^*)^{n(f_i - f_i')} = e^{n\zeta}$. We want to maximize this, equivalently, its log, and relaxing the requirement that $f$, $f'$ have rational entries we obtain the linear program (4.6).

**Proof of Proposition 4.9**

Since $\Sigma$ is a real symmetric matrix, it is positive-semidefinite and has a unique symmetric square root $\Sigma^{1/2} = U\,\Lambda^{1/2}\,U^T$, where $U$ is the matrix with the orthonormal eigenvectors of $\Sigma$ as columns (and $U$ is unitary, or real orthogonal), and $\Lambda$ is the diagonal matrix with the eigenvalues of $\Sigma$ on the diagonal. Further, $\Sigma^{1/2}$ is positive-definite, hence non-singular, if $\Sigma$ is ([HJ90], Theorem 7.2.6; also [BV04], §A.5.2). So it remains to show that $\Sigma$ is non-singular, i.e. that there is no vector $c \neq \mathbf{0}$ s.t. $c^T \Sigma\, c = 0$. If there were such a $c$, this would mean that the scalar r.v. $c^T \cdot \mathbf{u}$ is such that $\mathrm{var}(c^T \cdot \mathbf{u}) = c^T \Sigma\, c = 0$, so $c^T \cdot \mathbf{u}$ would be a constant with probability 1 ([Fel71], §III.5). Then the definition (4.8) of the $u_i$ would imply that $c_1 A'_{1,\cdot} + \cdots + c_\ell A'_{\ell,\cdot}$ is an $m$-vector with identical elements, so that the columns of $A'$ would be linearly-dependent, hence the rows would also be linearly dependent, contrary to our assumption in §4.3 and §1.

**Proof of Lemma 4.14**

We employ McDiarmid's concentration inequality, also known as "the method of bounded differences" ([McD98], Corollary 5.2 of [DP09], and [RS14]). Consider a sequence $x_1, \ldots, x_n$ of i.i.d. random variables taking values in $\{1, \ldots, m\}$ according to the p.d. $\varphi^*$. Let $f$ be the frequency vector of $x$ and define the function

$$g(x_1, \ldots, x_n) \triangleq \|f - \varphi^*\|_1.$$

This function satisfies a Lipschitz or bounded differences condition: if two sequences differ only in the $k$th element, their count vectors will differ by $\pm 1$ in exactly two elements, so

$$d_i \triangleq |g(x_1, \ldots, x_k, \ldots, x_n) - g(x_1, \ldots, x_k', \ldots, x_n)| = \tag{B.1}$$
$$|\,\|f - \varphi^*\|_1 - \|f' - \varphi^*\|_1\,| \leqslant \|(f - \varphi^*) - (f' - \varphi^*)\|_1 \leqslant 2/n.$$

Now McDiarmid's inequality states that for such a $g$,

$$\Pr(g > E(g) + t) \leqslant e^{-2t^2/d^2}, \qquad d^2 \triangleq \sum_i d_i^2. \tag{B.2}$$

From (B.1), $d^2 = 4/n$. The expectation of $g$ under $\varphi^*$ is not easy to calculate[B.1], but an upper bound will do. By the non-negativity of the variance, $E(|y|) \leqslant \sqrt{E(|y|^2)} = \sqrt{E(y^2)}$ for any $y$. Thus $E(|\nu_i - n\,\varphi_i^*|) \leqslant \sqrt{E((\nu_i - n\,\varphi_i^*)^2)} = \sqrt{n\,\varphi_i^*\,(1 - \varphi_i^*)}$ as the p.d. of $\nu$ is multinomial with parameter vector $\varphi^*$. So

$$E(g) = E(\|f - \varphi^*\|_1) = \frac{1}{n} \sum_{1 \leqslant i \leqslant m} E(|\nu_i - n\,\varphi_i^*|) \leqslant \frac{1}{\sqrt{n}} \sum_{1 \leqslant i \leqslant m} \sqrt{\varphi_i^*\,(1 - \varphi_i^*)} \triangleq \frac{1}{\sqrt{n}}\,s_1(\varphi^*). \tag{B.3}$$

---

[B.1]. The expectation of $|\nu_i - n\,\varphi_i^*|$ is the absolute first moment of the binomial p.d., and has a closed but unwieldy form.

Using (B.1) and (B.3) in (B.2), for any $t > 0$,

$$\mathrm{Pr}_{\varphi^*}(\|f - \varphi^*\|_1 > s_1(\varphi^*)/\sqrt{n} + t) \leqslant e^{-nt^2/2}.$$

The first result of the lemma then follows by setting $\vartheta = s_1(\varphi^*)/\sqrt{n} + t$.

Turning to the $\ell_2$ norm, take $g(x_1, ..., x_n) \triangleq \|f - \varphi^*\|_2$, so $d_i \leqslant \|(f - \varphi^*) - (f' - \varphi^*)\|_2 = \sqrt{2}/n$. Thus $d^2 = 2/n$ in this case. Next $E(g) \leqslant \sqrt{E(\|f - \varphi^*\|_2^2)}$, and, from what we did above, $E(\sum_i (f_i - \varphi_i^*)^2) = (1/n) \sum_{1 \leqslant i \leqslant m} \varphi_i^* (1 - \varphi_i^*) = (1 - s_2(\varphi^*))/n$, so $E(g) \leqslant \sqrt{1 - s_2}/\sqrt{n}$. The second result of the lemma then follows by setting $\vartheta = \sqrt{(1 - s_2)/n} + t$.

All of the above results still hold if we replace $\varphi^*$ by an arbitrary p.d. $\varphi$, so the lemma is valid for any p.d. $\varphi$, not just for a p.d. of maximum entropy.

# Appendix C  Auxiliary proofs

**Proof of (A.10)**

We give a somewhat lengthy proof of this inequality here, there may be a shorter one:

$$\sum_{\substack{\nu_1 + \cdots + \nu_\mu = n \\ \nu_1, ..., \nu_\mu \geqslant 1}} \frac{1}{\sqrt{\nu_1 \cdots \nu_\mu}} \leqslant \int_{\substack{x_1 + \cdots + x_\mu = n \\ x_1, ..., x_\mu \geqslant 0}} \frac{dx_1 \cdots dx_\mu}{\sqrt{x_1 \cdots x_\mu}} = \frac{\pi^{\mu/2}}{\Gamma(\mu/2)} n^{\mu/2 - 1}.$$

(See [GR80], 4.635, #4 for the integral.)

Consider the simplest case $\mu = 2$ first. We bound the sum as

$$\sum_{\substack{\nu_1 + \nu_2 = n \\ \nu_1, \nu_2 \geqslant 1}} \frac{1}{\sqrt{\nu_1 \nu_2}} = \sum_{\nu=1}^{n-1} \frac{1}{\sqrt{\nu(n-\nu)}} < \int_0^n \frac{dx}{\sqrt{x(n-x)}} = \pi. \tag{C.1}$$

This is because $\sum_{\nu=1}^{n/2} 1/\sqrt{\nu(n-\nu)} < \int_0^{n/2} dx/\sqrt{x(n-x)} = \pi/2$; the sum is a lower Riemann sum for the integral. Since the summand is symmetric about $n/2$, doubling the above produces the desired result.

Now consider the case of even $\mu$, i.e. $\mu = 2\lambda$. Divide the $\nu_i$ into $\lambda$ pairs, each of which sums to some number $\geqslant 2$ and these numbers in turn sum to $n$:

$$\sum_{\substack{\nu_1 + \cdots + \nu_{2\lambda} = n \\ \nu_1, ..., \nu_{2\lambda} \geqslant 1}} \frac{1}{\sqrt{\nu_1 \cdots \nu_{2\lambda}}} = \sum_{\substack{k_1 + \cdots + k_\lambda = n \\ k_1, ..., k_\lambda \geqslant 2}} \left( \sum_{\substack{\nu_1 + \nu_2 = k_1 \\ \nu_1, \nu_2 \geqslant 1}} \frac{1}{\sqrt{\nu_1 \nu_2}} \cdots \sum_{\substack{\nu_{2\lambda-1} + \nu_{2\lambda} = k_\lambda \\ \nu_{2\lambda-1}, \nu_{2\lambda} \geqslant 1}} \frac{1}{\sqrt{\nu_{2\lambda-1} \nu_{2\lambda}}} \right)$$

$$< \pi^\lambda \sum_{\substack{k_1 + \cdots + k_\lambda = n \\ k_1, ..., k_\lambda \geqslant 2}} 1.$$

Here the inequality follows by applying (C.1), which does not depend on $n$, to each of the inner sums. Further,

$$\sum_{\substack{k_1 + \cdots + k_\lambda = n \\ k_1, ..., k_\lambda \geqslant 2}} 1 = \sum_{\substack{k_1 + \cdots + k_\lambda = n - 2\lambda \\ k_1, ..., k_\lambda \geqslant 0}} 1 = \binom{n - \lambda - 1}{\lambda - 1},$$

where in the first equality we assume w.l.o.g. that $2\lambda < n$, and the 2nd equality follows from the fact that the number of compositions of $N$ into $M$ parts (i.e. the solutions of $k_1 + \cdots + k_M = N$, $k_i \geqslant 0$), is $\binom{N+M-1}{M-1}$. Finally we bound the binomial coefficient by $\binom{n-\lambda-1}{\lambda-1} < \frac{n^{\lambda-1}}{(\lambda-1)!}$, to arrive at

$$\sum_{\substack{\nu_1 + \cdots + \nu_{2\lambda} = n \\ \nu_1, \ldots, \nu_{2\lambda} \geqslant 1}} \frac{1}{\sqrt{\nu_1 \cdots \nu_{2\lambda}}} < \frac{\pi^\lambda}{\Gamma(\lambda)} \, n^{\lambda-1}. \tag{C.2}$$

Now we turn to the case of odd $\mu$, i.e. $\mu = 2\lambda + 1$. Similarly to what we did above,

$$\sum_{\substack{\nu_1 + \cdots + \nu_{2\lambda} + \nu_{2\lambda+1} = n \\ \nu_1, \ldots, \nu_{2\lambda}, \nu_{2\lambda+1} \geqslant 1}} \frac{1}{\sqrt{\nu_1 \cdots \nu_{2\lambda} \, \nu_{2\lambda+1}}} = \sum_{\substack{k_1 + k_2 = n \\ k_1 \geqslant 1, k_2 \geqslant 2\lambda}} \left( \sum_{\nu_{2\lambda+1} = k_1} \frac{1}{\sqrt{\nu_{2\lambda+1}}} \sum_{\substack{\nu_1 + \cdots + \nu_{2\lambda} = k_2 \\ \nu_1, \ldots, \nu_{2\lambda} \geqslant 1}} \frac{1}{\sqrt{\nu_1 \cdots \nu_{2\lambda}}} \right).$$

By (C.2), the r.h.s. does not exceed

$$\frac{\pi^\lambda}{\Gamma(\lambda)} \sum_{\substack{k_1 + k_2 = n \\ k_1 \geqslant 1, k_2 \geqslant 2\lambda}} \frac{k_2^{\lambda-1}}{\sqrt{k_1}} < \frac{\pi^\lambda}{\Gamma(\lambda)} \sum_{k=1}^{n-1} \frac{k^{\lambda-1}}{\sqrt{n-k}},$$

and this last sum can be bounded by the integral

$$\int_0^n \frac{k^{\lambda-1}}{\sqrt{n-k}} \, dk = n^{\lambda-1/2} \int_0^1 \frac{x^{\lambda-1}}{\sqrt{1-x}} \, dx = n^{\lambda-1/2} \frac{\Gamma(\lambda)\,\Gamma(1/2)}{\Gamma(\lambda+1/2)}.$$

We have thus shown that for $\mu = 2\lambda + 1$,

$$\sum_{\substack{\nu_1 + \cdots + \nu_{2\lambda+1} = n \\ \nu_1, \ldots, \nu_{2\lambda+1} \geqslant 1}} \frac{1}{\sqrt{\nu_1 \cdots \nu_{2\lambda+1}}} < \frac{\pi^{\lambda+1/2}}{\Gamma(\lambda+1/2)} \, n^{\lambda-1/2}. \tag{C.3}$$

Eqs. (C.2) and (C.3) establish (A.10) for all $\mu \geqslant 2$.